\documentclass[journal]{IEEEtran}
\IEEEoverridecommandlockouts
\usepackage{fixltx2e}
\usepackage{cite}
\usepackage{amsmath,amssymb,amsfonts}
\usepackage{graphicx}
\usepackage{textcomp}
\usepackage{xcolor}
\usepackage[ruled,linesnumbered]{algorithm2e}
\usepackage{amsthm}
\usepackage{comment}
\usepackage{url}
\usepackage{dsfont}
\usepackage{bm}
\usepackage{enumitem}
\usepackage{caption}
\usepackage[labelformat=simple]{subcaption}

\usepackage{xpatch}

\makeatletter
\ExplSyntaxOn
\cs_new:Npn \bibColoredItems #1#2
  {
    \clist_map_inline:nn {#2} { \cs_new:cpn {bib@colored@##1} {#1} }
  }
\ExplSyntaxOff

\newcommand\bib@setcolor[1]{%
  \ifcsname bib@colored@#1\endcsname
    \expandafter\color\expandafter{\csname bib@colored@#1\endcsname}
  \else
    \normalcolor
  \fi
}

\xpatchcmd\@bibitem
  {\item}
  {\bib@setcolor{#1}\item}
  {}{\PatchFailed}

\xpatchcmd\@lbibitem
  {\item}
  {\bib@setcolor{#2}\item}
  {}{\PatchFailed}
\makeatother

\def\BibTeX{{\rm B\kern-.05em{\sc i\kern-.025em b}\kern-.08em
    T\kern-.1667em\lower.7ex\hbox{E}\kern-.125emX}}

\begin{document}
\title{Adaptive Device-Edge Collaboration on DNN Inference in AIoT: A Digital Twin-Assisted Approach}
\author{Shisheng Hu, \IEEEmembership{Student Member, IEEE}, Mushu Li, \IEEEmembership{Member, IEEE}, Jie Gao, \IEEEmembership{Senior Member, IEEE}, \\Conghao Zhou, \IEEEmembership{Member, IEEE}, and Xuemin (Sherman) Shen, \IEEEmembership{Fellow, IEEE}\\
\thanks{Shisheng Hu, Conghao Zhou, and Xuemin (Sherman) Shen are with
the Department of Electrical and Computer Engineering, University of
Waterloo, Waterloo, ON, Canada, N2L 3G1 (email: \{shisheng.hu, c89zhou, sshen\}@uwaterloo.ca).}
\thanks{Mushu Li is with the Department of Electrical, Computer and Biomedical
Engineering, Toronto Metropolitan University, Toronto, Canada M5B 2K3
(email: mushu.li@ieee.org).}
\thanks{Jie Gao is with the School of Information Technology, Carleton University,
Ottawa, ON, Canada, K1S 5B6 (email: jie.gao6@carleton.ca).}
\thanks{Part of this paper was presented at the IEEE Global Communications Conference, Rio de Janeiro, Brazil, 2022 \cite{glopaper2022}. This work was supported by the Natural Sciences and Engineering Research Council (NSERC) of Canada. }
}

\maketitle
\begin{abstract}
Device-edge collaboration on deep neural network (DNN) inference is a promising approach to efficiently utilizing network resources for supporting artificial intelligence of things (AIoT) applications. In this paper, we propose a novel digital twin (DT)-assisted approach to device-edge collaboration on DNN inference that determines whether and when to stop local inference at a device and upload the intermediate results to complete the inference on an edge server. Instead of determining the collaboration for each DNN inference task only upon its generation, multi-step decision-making is performed during the on-device inference to adapt to the dynamic computing workload status at the device and the edge server. To enhance the adaptivity, a DT is constructed to evaluate all potential offloading decisions for each DNN inference task, which provides augmented training data for a machine learning-assisted decision-making algorithm. Then, another DT is constructed to estimate the inference status at the device to avoid frequently fetching the status information from the device, thus reducing the signaling overhead. We also derive necessary conditions for optimal offloading decisions to reduce the offloading decision space. Simulation results demonstrate the outstanding performance of our DT-assisted approach in terms of balancing the tradeoff among inference accuracy, delay, and energy consumption. 
\end{abstract}

\section{Introduction}
Artificial intelligence of things (AIoT) is receiving increasingly attention due to the remarkable capability of artificial intelligence (AI) on data analysis for Internet of Things (IoT) applications such as smart cities and smart manufacturing \cite{shen2021holistic, AIoTsurvey}.
Deep neural network (DNN) has been widely adopted in AIoT due to its ability of automatic feature extraction for pattern recognition or decision-making \cite{shen2021holistic, AIoTsurvey}. DNNs for AIoT applications are commonly deployed on cloud servers and edge servers since AIoT devices, e.g., smart cameras, have limited computing capabilities. Cloud servers, which possess extensive computing capabilities and data, can be leveraged for DNN training. In addition, edge servers, deployed in close proximity of AIoT devices, can process offloaded task data and generate processing results by DNN inference with low latency\cite{zhou2019edge}. With the growing popularity of AIoT applications, the demand for DNN inference tasks (referred to as DNN tasks) can be substantial, e.g., tens of frames are generated per second by a smart camera for real-time intersection monitoring \cite{huang2020intelligent}. In this regard, optimizing the performance of DNN inference via efficient utilization of network resources becomes a significant issue.

DNN partitioning is a promising solution to achieve the collaboration on DNN inference between AIoT devices and edge servers for enhanced performance of DNN inference. In DNN partitioning, the inference of a multi-layer DNN can be partitioned into two parts, i.e., on-device inference and edge inference\cite{kang2017neurosurgeon}. Specifically, an AIoT device executes the layers before a partition layer, i.e., on-device inference, and uploads the intermediate result, as the input to the partition layer, to an edge server. The edge server then executes the remaining layers of the DNN, i.e., edge inference, to obtain the final inference result. The partition layer for a DNN task determines computing load of on-device inference and edge inference, respectively, and the size of data to be offloaded. Therefore, the partition layer should be properly chosen to minimize the overall delay and energy consumption for task processing \cite{chen2021energy}. When the edge server is heavily loaded, the overall delay to process a task can be high, even if the optimal partition layer is chosen for the task. In this case, the DNN inference can be early exited by AIoT devices for some tasks instead of offloading them to the edge server for completing the inference\cite{distributed_TMC_confidence}. In this way, the overall delay of task processing can be reduced at the cost of reduced inference accuracy \cite{zeng2019boomerang}. By determining whether to offload a DNN task to the edge server or not, such a tradeoff can be made as needed.
With device-edge collaboration, DNN task offloading decisions are made on whether and when an AIoT device should stop on-device inference and upload intermediate results to an edge server to complete the inference. Such decision-making should adapt to the computing workload status at the device and the edge server, which is challenging since the workload status can change dynamically due to stochastic task generation at each AIoT device covered by the edge server \cite{impatient_queuing}. In particular, conventional approaches that make the offloading decision for each DNN task only upon its generation is unlikely to be optimal \cite{optimalStoppingDNN}. This is because on-device inference delay for a task can be as long as hundreds of milliseconds for executing one convolutional layer \cite{zeng2019boomerang}, which can result in \textit{i}) a drastic change of the instantaneous workload status between the generation of the task and the completion of the on-device inference for the task, 
and \textit{ii}) difficulty in predicting such a change especially when the dynamics of task arrival is unavailable \cite{zeng2019boomerang,taskstreamDNNinference}. 

The digital twin (DT) paradigm, proposed as a promising next stage of network virtualization \cite{shen2021holistic}, can be leveraged to facilitate adaptive device-edge collaboration on DNN inference. Through data synchronization, DTs can be constructed as digital representations of physical objects or processes to manage network data \cite{khan2022digital}, estimate network status (e.g., the running states of IoT devices \cite{IoT_DT_migration}), and predict network performance (e.g., the end-to-end delay of network slices\cite{DT_slice}). In addition, DTs can empower network emulations for efficient network management \cite{shen2021holistic}, e.g., constructing a simulation environment for pre-training a machine learning-assisted task offloading algorithm \cite{sun2020reducing}. For the case of device-edge collaborative DNN inference, DTs can empower machine learning-assisted offloading decision-making by providing augmented training data. For example, after a DNN task is offloaded to an edge server, DTs can be used to evaluate alternative offloading decisions of the task in hypothetical scenarios, i.e., offloading the task later or locally completing the task. 

In this paper, we leverage DTs to facilitate adaptive device-edge collaborative DNN inference. We consider an AIoT scenario, where a full-size DNN is deployed at an edge server for high-accuracy inference and a shallow DNN, which consists of the first several layers of the full-size DNN concatenated by an exit branch, is deployed at an AIoT device for low-latency on-device inference. In addition, the on-device and edge server computing workloads are dynamic due to stochastic task arrival. The objective is to optimize the performance of DNN inference in terms of balancing inference accuracy, delay, and energy consumption by determining the device-edge collaboration for the DNN tasks generated by the AIoT device. 
To achieve this objective, we propose a DT-assisted approach to adaptive device-edge collaboration on DNN inference, which determines whether to offload a DNN task each time when the on-device inference status changes, i.e., when a layer of the shallow DNN is about to be locally executed for the task. An offloading decision-making algorithm based on the optimal stopping theorem and assisted by machine learning is developed. 
Moreover, DTs are constructed in the proposed approach to play the following two roles:
\begin{enumerate}
\item Evaluating all potential offloading decisions of each DNN task to provide augmented training data for the learning-assisted offloading decision-making algorithm.
\item Estimating the inference status at the device to avoid frequently fetching the status information from the device, thus reducing the signaling overhead.
\end{enumerate}

The main contributions of this paper are summarized as follows:
\begin{enumerate}
\item We propose an approach to device-edge collaboration on DNN inference that adapts to the dynamic on-device and edge server workloads in offloading decision-making to optimize the DNN inference performance. 
\item We propose a learning-assisted algorithm, which generates effective offloading decisions with unknown task arrival statistics. Empowered by DTs, the algorithm can promptly attain an effective offloading solution with sufficient training data. 
\item We derive necessary conditions for optimal offloading decisions and  accordingly use them for decision space reduction, which reduces the complexity for adaptive device-edge collaboration on DNN inference. 
\end{enumerate}

The remainder of the paper is organized as follows. In Section \ref{related_work}, we review the related works. In Section \ref{system_model}, we introduce the system model. In Section \ref{DT_assisted_chap}, a DT-assisted approach to adaptive device-edge collaboration on DNN inference is proposed. The problem is formulated and transformed in Section \ref{problem_formulation_chap}, and the DT and learning-assisted offloading decision-making algorithm is proposed in Section \ref{DT-assited decision_chap}. In Section \ref{decison_reduction_chap}, we investigate offloading decision space reduction. Simulation results are provided in Section \ref{simulation_chap}. Section \ref{conclusion_chap} concludes the paper.


\section{Related Works}\label{related_work}
Collaboration among end devices and computing servers on task processing has attracted increasing research attention for efficient utilization of network computing resources. The collaboration among various computing servers within an IoT network was investigated in \cite{Mushu_collab_comp, conghao_TWC}. For the collaboration of AIoT devices and computing servers on DNN inference, some existing works optimized the number of DNN tasks offloaded from each AIoT device to the computing servers to minimize the overall delay in task completion \cite{9442308,IoT_hierarchy, fan2021accuracy}. 
Due to the limited computing capabilities, a lightweight DNN was executed by AIoT devices for task processing to reduce the delay of on-device DNN inference at the cost of inference accuracy \cite{fan2021accuracy, IoT_hierarchy, wu2020accuracy}. In such a case, the offloading decision was made to optimize the tradeoff between DNN inference costs (in terms of delay, energy consumption, etc.) and inference accuracy. 



Leveraging DNN task partitioning, some works investigated DNN task offloading by selecting the partition layer in light of server workload \cite{kang2017neurosurgeon, gao2021task, jointDNNmultiuser}. In \cite{kang2017neurosurgeon}, DNN partitioning-based device-edge collaboration was proposed, and the partition layer to minimize the delay or energy consumption of a DNN task was dynamically selected based on the real-time workload at a computing server. In a multi-device scenario, partition layer selection was jointly optimized with task scheduling and computing resource allocation at an edge server \cite{gao2021task,jointDNNmultiuser}. Other works evaluated the on-device computing workload for DNN task partitioning\cite{throughput_infocom_trans, taskstreamDNNinference}. Specifically, considering periodical task generation at the device,  Hu \textit{et al.} maximized the throughput of the DNN inference while preventing the DNN tasks from experiencing on-device queuing delay \cite{throughput_infocom_trans}. For the case when the DNN task generation at the device is non-periodical, Song \textit{et al.} proposed a deep reinforcement learning (DRL)-based algorithm to dynamically select the partition layer \cite{taskstreamDNNinference}.
The aforementioned works adapted to the network dynamics, e.g., dynamic channel condition \cite{TII_DNN}, on-device workload \cite{taskstreamDNNinference}, and server workload\cite{kang2017neurosurgeon}, by making offloading decision for each DNN task upon its generation. In order to make the device-edge collaboration more adaptive, the offloading decision of each DNN task was adjusted during the on-device DNN inference based on the real-time estimation and the statistics of channel conditions \cite{optimalStoppingDNN}.

Different from the existing works, this work proposes a DT-assisted approach to device-edge collaboration on DNN inference that can adapt to the dynamic on-device and edge server computing workloads with unknown task arrival statistics. In addition, we investigate how to reduce the signaling overhead for and the complexity of adaptive device-edge collaboration on DNN inference, by establishing a DT and by analyzing properties of the workload evolution, respectively. Moreover, instead of simply employing a learning algorithm to identify unknown network dynamics in offloading decision-making, we investigate how a DT can be used to augment the training data for empowering such a learning algorithm. \looseness=-1


\section{System Model}\label{system_model}
In this section, we first introduce the models of DNN task generation, computing and queuing. Then, we define the utility of a DNN task, which incorporates the delay, inference accuracy, and energy consumption for processing a DNN task. 

\subsection{Task Generation Model}
Consider an AIoT device that connects to an edge computing server through an access point (AP). 
The AIoT device collects sensing data, such as images, and generates computing tasks such as object recognition for processing the collected data. 
The time horizon is divided into slots with equal duration denoted by $\Delta T$, and the index of a time slot is denoted by $t\in\{1,2,...\}$. At the beginning of each time slot, a task is probabilistically generated by the device with an unknown probability \cite{taskstreamDNNinference}. Let $\Delta T_n$ represent the interval between the time instants when the ($n+1$)-th and the $n$-th tasks are generated, where $n \in \{1,2,...,N\}$, and let $\Delta T_0$ represent the time instant when the first task is generated.
\subsection{Task Computing Model}
\begin{figure}
    \centering
    \includegraphics[width=8.0cm]{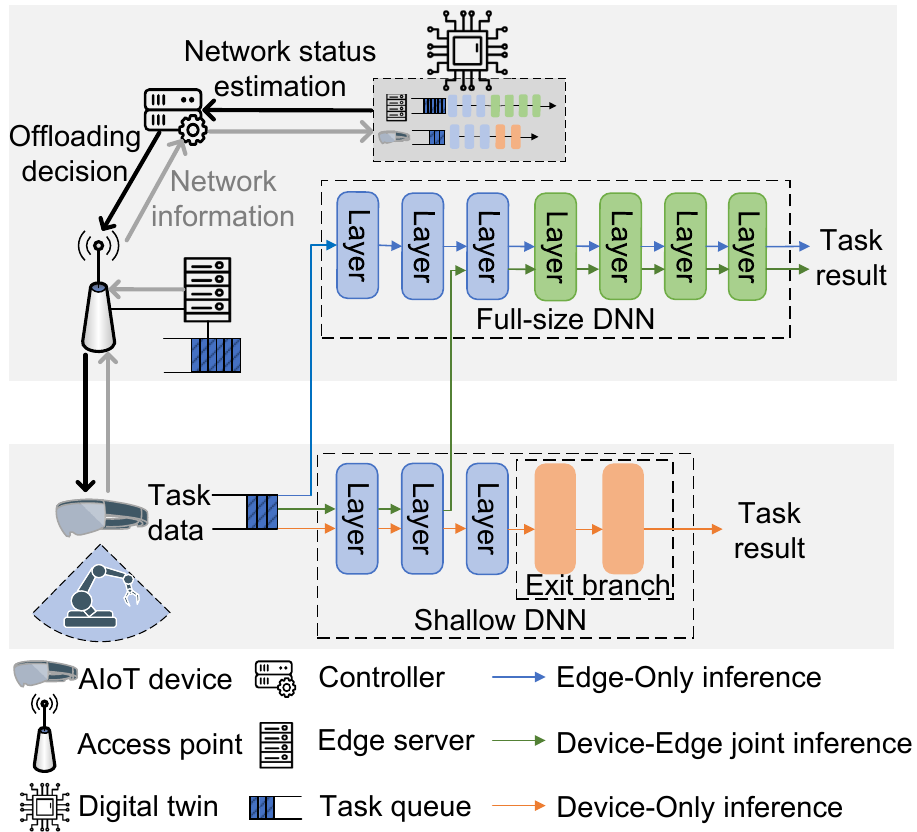}
    \caption{System model.}
    \label{system}
\end{figure}

For processing the tasks, one full-size DNN with $L$ consecutive layers (e.g., $L=7$ in Fig. \ref{system}) is deployed at the edge server, and one shallow DNN with fewer layers is deployed at the AIoT device. For DNNs that contain parallel layers or residual blocks, the layers in each residual block or the layers with parallel execution can be abstracted as one logical layer\cite{Batching_TWC_inference, jointDNNmultiuser}. In the design and training of the full-size and shallow DNNs, the BranchyNet architecture \cite{teerapittayanon2016branchynet} is used such that the first $l_e$ layers of the two DNNs are identical (e.g., the three blue-colored layers in Fig. \ref{system}). The remaining part of the shallow DNN is referred to as an exit branch, and the layers therein (e.g., the two orange-colored layers in Fig. \ref{system}) are abstracted as one logical layer, i.e., the ($l_e+1$)-th layer of the shallow DNN.

Each time before the AIoT device executes the next layer in the shallow DNN, a network controller decides whether the AIoT device should stop locally processing the task at this point and offload the task to the edge server instead. For the $n$-th task, when the task is offloaded to the edge server or locally completed, the number of executed layers in the shallow DNN for the task is determined, which is denoted by $x_n \in \{0,1,...,l_e+1\}$. We define $x_n$ as the offloading decision for the $n$-th task, which can result in three DNN inference scenarios as illustrated in Fig. \ref{system}:
\begin{enumerate}
\item \textit{Edge-Only Inference:} If $x_n=0$, the $n$-th DNN task is offloaded to the edge server without being processed by any layer of the shallow DNN by the device. In other words, the task is completely processed by the full-size DNN at the edge server.\looseness=-1 
\item \textit{Device-Edge Joint Inference:} 
If $1\leqslant x_n\leqslant l_e$, the $n$-th DNN task is locally processed by the first $x_n$ layers of the shallow DNN (e.g., $x_n=2$ for the device-edge joint inference scenario in Fig. \ref{system}). Then, the output of the $x_n$-th layer of the shallow DNN, as the intermediate result, is uploaded to the edge server and processed by the remaining layers of the full-size DNN. 
\item \textit{Device-Only Inference:} If $x_n=l_e+1$, the $n$-th DNN task is not offloaded to the edge server but completely processed by the shallow DNN at the device. 
\end{enumerate}

To assist in making offloading decisions of DNN tasks, as shown in Fig. \ref{system}, DTs are constructed by the network controller. Specifically, the network controller collects the network information, i.e., the per-layer on-device DNN inference delay and the information on task arrivals at the AIoT device and the edge server. The collected information is processed by DT models to estimate the status of the AIoT device and the edge server for assisting in making offloading decisions of DNN tasks. Detailed introduction of the DTs will be presented in Section \ref{DT_assisted_chap}.

\begin{figure}
    \centering
    \includegraphics[width=8.0cm]{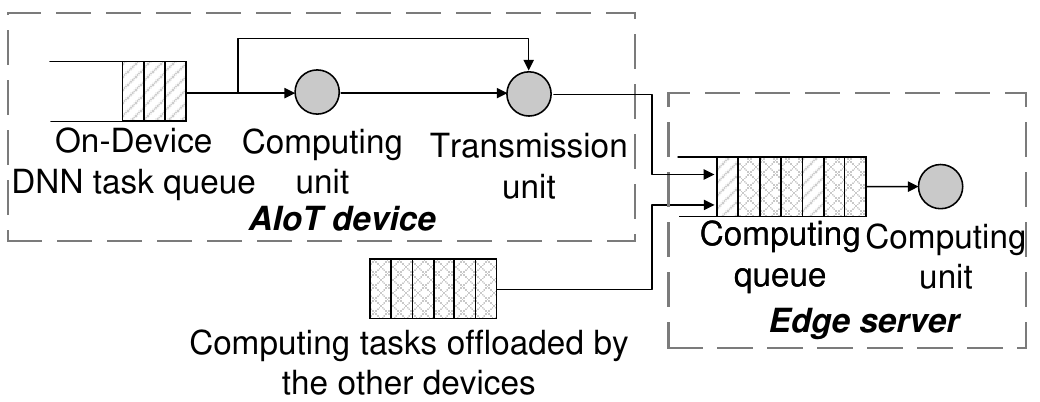}
    \caption{Task queuing model.}
    \label{scheduling}
\end{figure}

\subsection{Task Queuing Model}
As shown in Fig. \ref{scheduling}, the task queuing model includes queuing at the device and queuing at the edge server.

\textit{1) On-Device Task Queue:} The device has one computing unit and one transmission unit, which can respectively process and transmit only one task at any time instant. The tasks that have been generated but not yet processed or offloaded are stored in an on-device task queue following the first-come-first-serve (FCFS) rule. 
When the computing unit becomes idle, the first task in the on-device task queue leaves the queue. If the transmission unit is also idle, edge-only inference, i.e., the transmission unit offloads the task to the edge server without any on-device processing, can be chosen for the task. Otherwise, the task will be processed by the computing unit until the transmission unit is idle and a decision to offload the task is made. At the beginning of the ($t\!+\!1$)-th time slot, the on-device computing workload in terms of the number of tasks in the on-device queue, denoted by $Q^{D}(t\!+\!1)$, is updated as:\looseness=-1
\begin{equation}\label{device_queue_evo}
Q^{D}(t\!+\!1)\!=Q^{D}(t)\!+\! I(t\!+\!1)\!-\! O(t\!+\!\!1),
\end{equation}
where $I(t)$ equals to $1$ if one DNN task is generated in the ($t-1$)-th time slot and $0$ otherwise; $O(t)$ equals to $1$ if a DNN task leaves the on-device task queue at the beginning of the $t$-th time slot and $0$ otherwise.

\textit{2) Edge Server Queue:} The edge server has one computing unit, which can process only one task at any time instant, and maintains a computing queue for tasks yet to be processed. The computing tasks arriving at the edge server in each time slot will either start to be processed by the computing unit or enter the computing queue at the beginning of the next time slot.\footnote{The processing order of the computing tasks arriving at the edge server in a time slot is determined by a task scheduling algorithm. For simplicity, we assume that the task offloaded by the considered device will be first processed.} The edge server workload is the sum of CPU cycles required to complete the processing of the task in the computing unit as well as all the tasks in the computing queue. At the beginning of the ($t\!+\!1$)-th time slot, the edge server workload, denoted by $Q^{E}(t\!+\!1)$, is updated as:
\begin{equation}\label{edge_queue_evo}
Q^{E}(t\!+\!1)=\max\{ Q^{E}(t)\!-\! f^E\Delta T,0 \}\!+\! D(t)\!+\!W(t),
\end{equation}
where $D(t)$ and $W(t)$ represent the workload of the tasks from the considered device and other devices connected to the edge server, respectively, in the $t$-th time slot; $f^E$ represents the computation frequency of the edge server in the unit of CPU cycles per second.


\subsection{Task Utility}
In this subsection, we first derive the delay, inference accuracy, and energy consumption of processing a DNN task. Then, we define the utility of a DNN task.

\textit{1) Delay:} Depending on the task offloading decision, the processing of a DNN task can incur \textit{i)} on-device queuing delay, \textit{ii)} on-device inference delay, \textit{iii)} transmission delay from uploading the intermediate result, \textit{iv)} queuing delay at the edge server, and \textit{v)} edge inference delay. The delay from delivering the task result to the device is neglected due to the typically small size of the result.

\textit{i) On-Device Inference Delay:} 
The on-device inference delay of the $n$-th task is the sum of the execution delay of the first $x_n$ layers in the shallow DNN, given by:
\begin{equation}\label{lc_delay}
T^{\rm lc}_n(x_n)=\left\{
\begin{aligned}
&\begin{matrix}\sum_{l=1}^{x_n}d_{l}^{D}\end{matrix}, x_n\ge1,
\\&0, x_n=0,
\end{aligned}
\right.
\end{equation}
where $d_{l}^{D}$ represents the time to execute the $l$-th layer in the shallow DNN by the device, $\forall l \in\{1,2,...,l_e+1\}$. The value of $d_{l}^{D}$ is rounded to an integer multiple of the time slot duration $\Delta T$.

\textit{ii) On-Device Queuing Delay:} The queuing delay of the $n$-th DNN task at the device is determined by the offloading decisions for the DNN tasks generated earlier, which are denoted as $\mathbf{x}_{n-1}=\{x_{1},x_{2},..., x_{n-1}\}$. Let $\mathbf{x}_{0}=\emptyset$ for consistency. The on-device queuing delay for the $n$-th task, denoted by $T^{\rm lq}_n(\mathbf{x}_{n-1})$, can be calculated by:
\begin{equation}\label{local_queuing}
\begin{split}
&T^{\rm lq}_n(\mathbf{x}_{n-1})\\
&= \left\{
\begin{aligned}
&0, ~n=1,
\\&\begin{matrix}\!\max\{T^{\rm lq}_{n-1}\!(\mathbf{x}_{n-2})\!+\!T^{\rm lc}_{n-1}(x_{n-1})\!-\!\Delta T_{n-1}, 0\}\end{matrix}, \!~n \!\ge\! 2.
\end{aligned}
\right.
\end{split}
\end{equation}

\textit{iii) Data Uploading Delay:} If edge-only inference or device-edge joint inference is adopted for the $n$-th DNN task, i.e., $0\le x_n\le l_e$, the task processing will incur an uploading delay denoted by $T_n^{\rm up}$. The data to be uploaded is the input to the $(x_n+1)$-th layer of the shallow DNN, and the uploading delay can be calculated by:
\begin{equation}
T_n^{\rm up}(x_n)= \left\{
\begin{aligned}
&s_{x_n}/R_0, 0 \le x_n\le l_e,
\\ &0, ~x_n=l_e+1,
\end{aligned}
\right.
\end{equation}
where $s_{l}$ represents the size of the data to the ($l+1$)-th layer of the shallow DNN, $\forall l \in\{0,1,...,l_e\}$; $R_0$ is the uplink transmission rate between the device and the AP. 

\textit{iv) Edge Queuing Delay:} If the $n$-th DNN task is offloaded to the edge server, i.e., $x_n\le l_e$, the task processing can incur a queuing delay at the edge server, given by:
\begin{equation}\label{edge_queuing}
T^{\rm eq}_n(\mathbf{x}_n)= \left\{
\begin{aligned}
&Q^{E}(t_{n,x_n})/f_s^E, ~0\le x_n\le l_e,
\\&0, ~ x_n=l_e+1,
\end{aligned}
\right.
\end{equation}
where $t_{n,x_n}$ is the index of the time slot when the $n$-th DNN task arrives at the edge server.\footnote{Assuming a high data transmission rate, the data uploading for a DNN task can be completed within one time slot.} 

\textit{v) Edge Inference Delay:} If $x_n\le l_e$, the task processing will incur an inference delay at the edge server from executing the remaining layers of the full-size DNN, given by:
\begin{equation}\label{ec_delay}
 T^{\rm ec}_n(x_n) = \left\{
\begin{aligned}
&\begin{matrix}\sum_{l=x_n+1}^{L} {d_{l}^{E}}\end{matrix}, ~0 \le x_n\le l_e,
\\ &0, ~ x_n=l_e+1,
\end{aligned}
\right.
 \end{equation}
where $d_{l}^{E}$ represents the execution time in the $l$-th layer of the full-size DNN by the edge server.
Summarizing the delay in equations (\ref{lc_delay}) to (\ref{ec_delay}), the overall delay of the $n$-th DNN task, denoted by $T_n(\mathbf{x}_n)$, can be calculated by:
\begin{equation}\label{queuing_delay}
T_n(\mathbf{x}_n) = T_n^{\rm lq}(\mathbf{x}_{n-1})+T^{\rm lc}_n(x_n) +T_n^{\rm up}(x_n)+T_n^{\rm eq}(\mathbf{x}_{n})+ T_n^{\rm ec}(x_n).
\end{equation}





\textit{2) Inference Accuracy:} The accuracy of the task result is denoted by $A_n(x_n)$, which equals $\eta^E$ if the task is offloaded to the edge server for processing by the full-size DNN, i.e., $1\!\leq\! x_n \!\leq\! l_e$, and $\eta^D$ if the result is derived by the shallow DNN, i.e., $x_n = l_e+1$. The accuracy of the task result derived by the full-size DNN is higher than that derived by the shallow DNN, i.e., $\eta^E\!>\! \eta^D$.

\textit{3) Energy Consumption:} The energy consumption for processing the $n$-th DNN task, denoted by $E_n(x_n)$, includes two parts, i.e., the energy consumption for DNN inference and that for task data uploading. The overall energy consumption can be calculated by:
\begin{equation}
E_n(x_n) =
\kappa^{D} (f_s^D)^3 T_n^{\rm lc} + \kappa^E(f_s^E)^3 T_n^{\rm ec}+ p^{\rm up}T_n^{\rm up},
\end{equation}
where $\kappa^D$ and $\kappa^E$ are the energy efficiency coefficients of the device and the edge server, respectively \cite{9442308}; $p^{\rm up}$ is the transmit power of the device.
%

\textit{4) Task Utility:} We define the utility of the $n$-th DNN task, denoted by $U_n(\mathbf{x}_n)$, as a weighted sum of the overall delay, inference accuracy, and energy consumption of the task: 
\begin{equation}\label{utility}
U_n(\mathbf{x}_n) = -T_n(\mathbf{x}_n)+\alpha A_n(x_n)-\beta E_n(x_n),
\end{equation}
where $\alpha$ and $\beta$ are positive weights for the inference accuracy and energy consumption, respectively.

\section{DT-Assisted Adaptive Device-Edge Collaboration on DNN Inference}\label{DT_assisted_chap}
In this section, we first introduce the data required to establish the DTs of two processes, i.e., a DT of the on-device DNN inference and a DT of the computing workload evolution. Then, we introduce how the DT of on-device inference can be established to estimate the on-device inference status, and how the DT of computing workload evolution can be established to emulate the on-device workload and edge server workload, respectively. Finally, assisted by the two DTs, an approach to adaptive device-edge collaboration on DNN inference is proposed.

\subsection{Data for Establishing DTs}\label{DT_data}
The following information for establishing the DTs is estimated or collected by the network controller:
\begin{itemize}
\item \textit{Per-Layer On-Device Inference Delay:} The delay for executing each layer of the shallow DNN by the device, i.e., $\{d_l^D, l=1,2,...,l_e+1\}$, is estimated by the network controller. Specifically, the estimation can be based on \textit{i)} the number of FLOPs required for each layer and the computing capability of the device \cite{FLops_calculation_JSAC}; or \textit{ii)} regression models given the configuration of each layer such as the input and output data sizes \cite{gao2021task}. 

\item \textit{Task Generation and Arrival:} At the beginning of each time slot, the controller collects information on \textit{i)} whether a DNN task is generated by the AIoT device at the beginning of each time slot, i.e., $\{I(t), t=1,2,...\}$; \textit{ii)} the workload of computing tasks arriving at the edge server not from the considered device in the time slot, i.e., $\{W(t), t=1,2,...\}$.
\end{itemize}

\subsection{DT of On-Device DNN Inference}
For each DNN task, the network controller determines whether to continue the on-device inference before the device executes each layer in the shallow DNN. As a result, the change of on-device DNN inference status, i.e., a layer is about to be locally executed, needs to be known by the network controller. Instead of acquiring the on-device inference status from the device in real-time, which can result in large signaling overhead, the network controller establishes a DT to emulate the on-device inference and estimate the on-device inference status. Denote the index of the time slot right before the on-device execution of the $(l+1)$-th layer for the $n$-th DNN task by $t_{n,l}, \forall l\in\{0,1,..., l_e\}$. Then, $t_{n,l}$ can be estimated based on the time instant when the $n$-th DNN task is generated, the on-device queuing delay of the $n$-th DNN task, and the on-device inference delay before executing the ($l+1$)-th layer:
\begin{equation}\label{exeuction_time_est}
t_{n,l}(\mathbf{x}_{n-1}) = \frac{1}{\Delta T}\Big(\big(\sum_{i=0}^{n-1}\Delta T_i\big)+T_n^{\rm lq}(\mathbf{x}_{n-1})+\sum_{i=0}^{l}d_i^{D}\Big).
\end{equation}
In addition, $t_{n,l_e+1}(\mathbf{x}_{n-1})$ is also calculated, which represents the index of the next time slot if the task were completed by device-only inference.


\subsection{DT of Computing Workload Evolution}
The evolution of the computing workload on the AIoT device and the edge server affects the average utility of the DNN tasks. To capture the evolution with unknown statistics, a learning algorithm can be applied when determining whether to continue the on-device inference for a DNN task. Such an algorithm can be trained based on the utility of the tasks and the decisions made for the tasks. However, if a decision to offload a task is made, options of continuing the on-device inference for the task cannot be evaluated for training the algorithm. This is because offloading the task would affect the workload evolution, which makes the workload evolution if the on-device inference continued unobservable.

To estimate the on-device and edge server computing workload status under each candidate offloading decision for a DNN task, the DT of computing workload evolution is established. Specifically, for each task, the DT emulates the workload evolution in a hypothetical case, i.e., if the task processing were completed by the shallow DNN at the device.
The time slots to execute the shallow DNN for the $n$-th task are denoted by $t \in \{t_{n,0},t_{n,0}+1,...,t_{n,l_e+1}-1\}$, 
and the emulated on-device workload and edge server workload, denoted by $\tilde{Q}^{D}(t)$ and $\tilde{Q}^{E}(t)$, respectively, are calculated by:
\begin{subequations}\label{dt_emulation}
\begin{equation}\label{emu_devicequeue}
\tilde{Q}^{D}(t)\!\!=\!\!\left\{\!
\begin{aligned}
& {Q}^{D}(t), t=t_{n,0},
\\&\!\tilde{Q}^{D}(t\!-\!1)\!+\! I(t), \rm{otherwise},
\end{aligned}
\right.
\end{equation}
\begin{equation}\label{emu_edgequeue}
\tilde{Q}^{E}(t)\!\!=\!\!\left\{\!
\begin{aligned}
& {Q}^{E}(t), t=t_{n,0},
\\&\!\max\{\tilde{Q}^{E}\!(t\!-\!1)-\!\!f^E\Delta T,0 \}\!\!+\!\! W(t), \rm{otherwise}.
\end{aligned}
\right.
\end{equation}
\end{subequations}

The difference between the actual computing workloads in (\ref{device_queue_evo}) and (\ref{edge_queue_evo}) and the emulated computing workloads in (\ref{dt_emulation}) is that the former are calculated based on the actual offloading decision of a DNN task while the latter are calculated in the hypothetical case that the task processing were locally completed. In such a hypothetical case, the decrease of the computation workload at the device due to the offloading of the task, i.e., $O(t)$ in (\ref{device_queue_evo}), and the increase of the computation workload at the edge server due to the offloading of the task, i.e., $D(t)$ in (\ref{edge_queue_evo}), both equal zero. Correspondingly, equation (\ref{dt_emulation}) for calculating the emulated workloads in the hypothetical case omits $O(t)$ and $D(t)$, which is the major difference with (\ref{device_queue_evo}) and (\ref{edge_queue_evo}).

\subsection{Analysis of DTs' Overheads}
The data collection and processing for DTs can result in a delay and resource consumption. Nevertheless, such a delay and resource consumption are relatively low. This is because, first, the data collected for DTs as described in subsection \ref{DT_data} have a small size. Second, the processing of the collected data, following (\ref{exeuction_time_est}) and (\ref{dt_emulation}), only involves several addition operations. Third, only the recently collected data, i.e., the information on tasks that arrive at the AIoT device and the edge server during the on-device processing of the current DNN task, are required for data processing. As a result, the outdated data, such as the information on tasks that arrive during the on-device processing of the preceding DNN tasks, can be discarded to minimize the cost of storing the DTs.

\subsection{DT-Assisted Approach to Adaptive Device-Edge Collaboration}
\begin{figure}
    \centering
    \includegraphics[width=8cm]{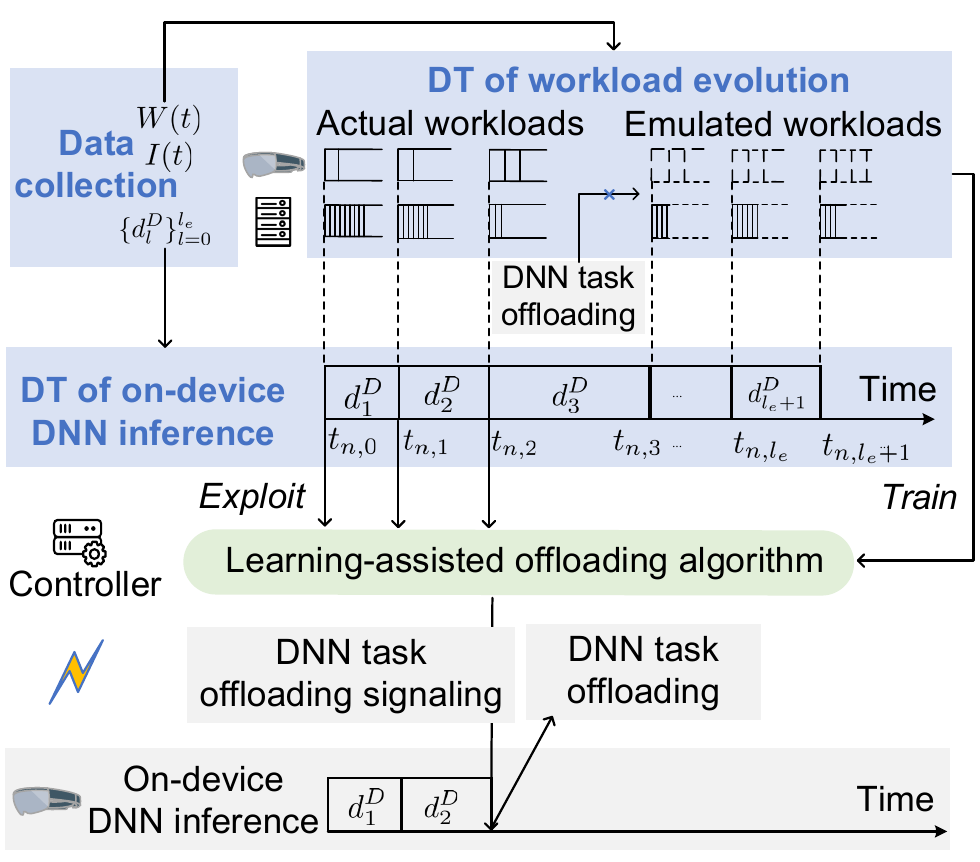}
    \caption{DT-Assisted approach to adaptive device-edge collaboration on DNN inference.}
    \label{DToverview}
\end{figure}

As shown in Fig. \ref{DToverview}, with the DT of on-device inference and the DT of computing workload evolution, we propose an approach to adaptive device-edge collaboration on DNN inference, which involves the following four steps.
\begin{enumerate}
  \item \textit{Step 1 (Task Information Gathering):} 
      In the beginning of each time slot, the device sends the DNN task generation indicator $I(t)$ to the controller, which then calculates the gaps of task generations $\{\Delta T_n,n=0,1,...,(\begin{matrix}\sum_t I(t))\end{matrix}\!-\!1\}$. After the controller determines the offloading decision for the ($n\!-\!1$)-th DNN task, i.e., $x_{n-1}$, the on-device queuing delay of the $n$-th DNN task, i.e., $T_n^{\rm lq}(\mathbf{x}_{n-1})$, is calculated. Then, the controller utilizes the DT of on-device inference and (\ref{exeuction_time_est}) to estimate the indices of the time slots when the device would execute each layer of the shallow DNN for the $n$-th task, i.e., $\{t_{n,l}(\mathbf{x}_{n})\}_{l=0}^{l_e}$, assuming the task continues to be processed locally.
  \item \textit{Step 2 (Learning-Assisted Offloading Decision-Making):} After the transmission unit is idle, i.e., $t\ge{t_{n,\widehat{x}_n}}$, at the beginning of any time slot $t\in\{t_{n,l}(\mathbf{x}_{n-1})\}_{l=\widehat{x}_n}^{l_e}$, leveraging a learning-assisted algorithm, the controller determines whether the device should continue the on-device execution of the $(l+1)$-th layer of the shallow DNN or upload the intermediate result to the edge server. 
  \item \textit{Step 3 (Signaling of Task Offloading):}  If the controller determines to stop the on-device inference for the $n$-th DNN task in the beginning of the $t$-th time slot, it sends a signal to the device, corresponding to setting $O(t)=1$, to the device. The device then uploads the intermediate result to the edge server for executing the remaining layers of the full-size DNN. Without receiving this signal, the device will continue executing the next layer of the shallow DNN.
  \item \textit{Step 4 (Training of Learning-Assisted Offloading Algorithm):} After the time slot $t\!=\!t_{n,l_e+1}(\mathbf{x}_{n-1})$, the workload evolution if the $n$-th DNN task were locally completed by the shallow DNN, is generated using the DT of computing workload evolution and (\ref{dt_emulation}). Then, a learning-assisted offloading algorithm will be trained with the data augmented by the emulated workload evolution.
\end{enumerate}

\section{Problem Formulation and Transformation}\label{problem_formulation_chap}

In this section, we formulate the problem of maximizing the average utility of DNN tasks generated by the considered AIoT device. Due to the stochastic DNN task generation, we transform the problem into multiple sub-problems, which are solved sequentially to make the offloading decision for each task given the decisions for its preceding tasks.

\subsection{Problem Formulation}
We aim to maximize the average utility of DNN tasks generated at the considered device by optimizing the offloading decisions for the tasks. The problem is given by:
\begin{subequations}\label{original_prob}
\begin{equation}\label{obeject}
{\rm(P1)}\bm{\max}_{\mathbf{x}_N} \lim_{N \to \infty}\frac{1}{N}\begin{matrix} \sum_{n=1}^{N} {U_n(\mathbf{x}_n)}\end{matrix}
\end{equation}
\begin{equation}\label{constraint1}
{\bm{\mathrm{s.t.}}} ~0\le x_n \le l_e+1, \forall n \in\{0,1,...,N\},
\end{equation}
\begin{equation}\label{constraint2}
\begin{split}
(\begin{matrix}\sum_{i=m}^{n-1}\end{matrix}\Delta T_{i})+ T_{n}^{\rm lq}(\mathbf{x}_{n-1})+T_n^{\rm lc}(x_n)\ge
T_{m}^{\rm lq}(\mathbf{x}_{m-1})~~~~~\\+T_{m}^{\rm lc}(x_m)+T_{m}^{\rm up}(x_m), \forall m \in \{1 ,...,n-1\}, n \in\{1,2,...,N\},
\end{split}
\end{equation}
\end{subequations}
where (\ref{constraint2}) ensures that the $n$-th DNN task can be offloaded
to the edge server only when the offloading of preceding DNN tasks, if any, has been completed and the transmission unit is idle. Note that $T_n^{\rm lc}(x_n)$ monotonically increases with $x_n$. As a result, given any $\mathbf{x}_{n-1}$, we can find for the $n$-th task a minimum number of layers in the shallow DNN, denoted by $\widehat{x}_n(\mathbf{x}_{n-1})$, which should be locally executed to satisfy (\ref{constraint2}). 
An equivalent form of constraints (\ref{constraint1}) and (\ref{constraint2}) is:
\begin{equation}\label{new_constraint}
\widehat{x}_{n}(\mathbf{x}_{n-1})\leq x_n\leq l_e+1, \forall n \in\{0,1,2...,N\}.
\end{equation}
\subsection{Problem Transformation}

Solving ${\rm(P1)}$ requires the joint optimization of offloading decisions for all the DNN tasks, which is challenging due to the stochastic task generation at the device with unknown statistics \cite{taskstreamDNNinference}. In this subsection, we first define the long-term utility of one task, which incorporates the main impact of the offloading decision of the task on the other tasks, i.e., the on-device queuing delay of the other tasks resulting from the on-device inference of the task.\footnote{Although the edge queuing delay of a DNN task from the considered device can be affected by the offloading decisions of its preceding DNN tasks, such effect is relatively small as the edge queuing delay is affected by the computing tasks offloaded to the edge server from the other devices.} Then, we transform (P1) into the online optimization of the offloading decision for each DNN task given the offloading decisions of the preceding DNN tasks, with the objective to maximize the expected long-term utility of each DNN task.

\begin{figure}
    \centering
    \includegraphics[width=8cm]{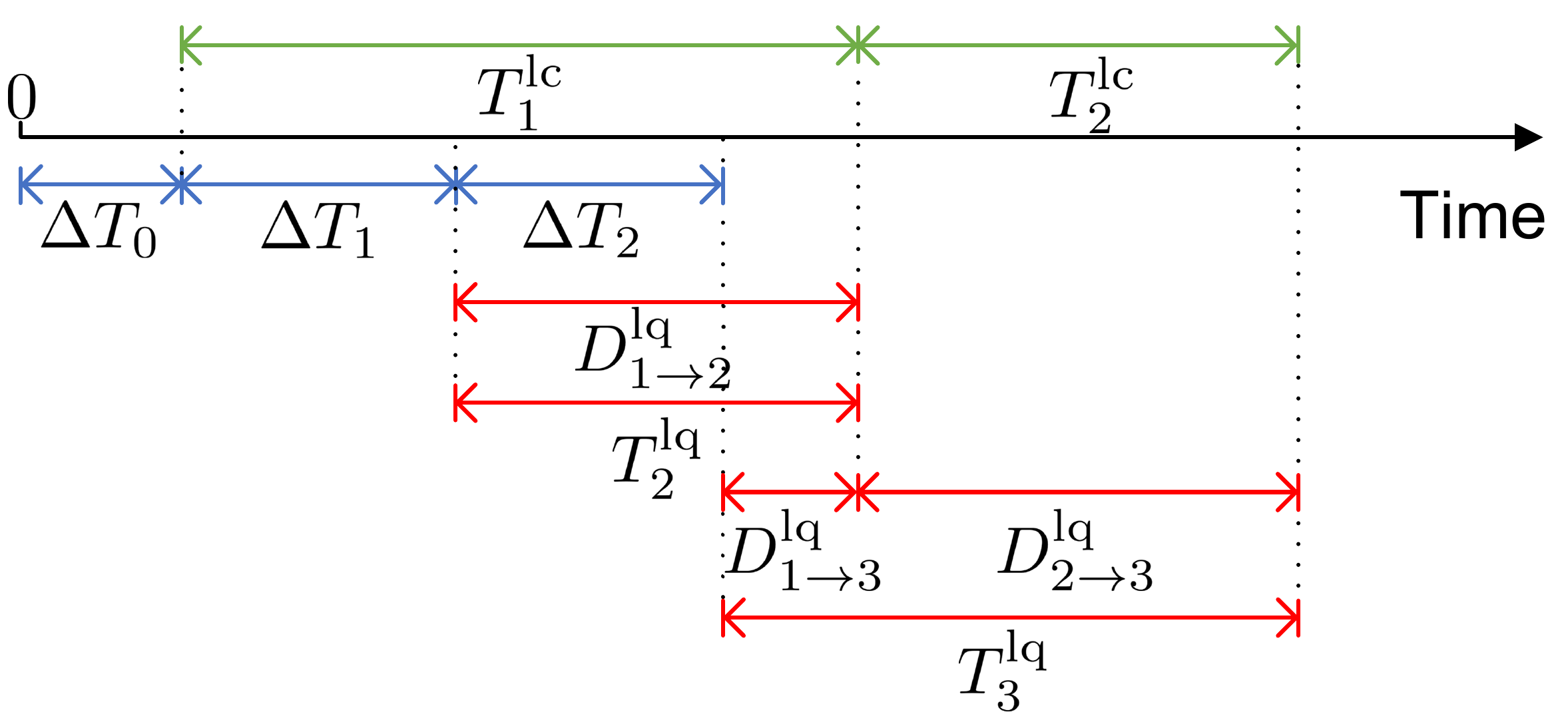}
    \caption{Decomposition of the on-device queuing delay $T_n^{\rm lq}$.}
    \label{queuing_delay_illustration}
\end{figure}

\textit{1) Long-Term Utility of DNN Task:} 
The on-device queuing delay of a DNN task results from the on-device inference for its preceding tasks. 
As shown by the red lines in Fig. \ref{queuing_delay_illustration}, we first decompose the on-device queuing delay of a task as the summation of the queuing delay resulting from each of its preceding tasks. Then, we define the long-term on-device queuing delay for processing each task, which is the sum on-device queuing delay of subsequent tasks resulting from the on-device processing of the task. For brevity, we omit the parentheses and the offloading decision variables in notations.\looseness=-1

First, define $D_{m\to n}^{\rm lq}$ as the on-device queuing delay of the $n$-th DNN task directly resulting from the on-device inference of the $m$-th DNN task, calculated by:
\begin{equation}\label{queuing_delay_to}
D_{m\to n}^{\rm lq}=\left\{
\begin{aligned}
& 0, ~ m\ge n,
\\&\max\{\min{\{T^{\rm lq}_{m}\!-\!\begin{matrix}\sum_{i=m}^{n\!-\!1}\!\Delta T_i\end{matrix}, 0\}} \!+ \!T_{m}^{\rm lc},0\}, ~ m\!<\! n.
\end{aligned}
\right.
\end{equation}
When $m\!\ge\! n$, $D_{m\to n}^{\rm lq}$ equals to zero since tasks leave the on-device task queue following the FCFS rule; when $m\!<\! n$, the maximum $D_{m\to n}^{\rm lq}$ is the on-device inference delay of the $m$-th task, i.e., $T_m^{\rm lc}$ (e.g., $D_{2\to 3}^{\rm lq}=T_2^{\rm lc}$ in Fig. \ref{queuing_delay_illustration}). Given the offloading decision of the $m$-th DNN task, $D_{m\to n}^{\rm lq}$ decreases with the increase of the gap between the generation instants of the $m$-th and the $n$-th tasks (e.g., $D_{1\to 3}^{\rm lq}$ decreases with the increase of $\Delta T_1$ or $\Delta T_2$ in Fig. \ref{queuing_delay_illustration}).

\textbf{\textit{Proposition 1:}} The on-device queuing delay of the $n$-th DNN task is the sum of the on-device queuing delay resulting from all the $N$ DNN tasks generated by the device:
\begin{equation}
T^{\rm lq}_{n}=\begin{matrix}\sum_{m=1}^{N} D_{m\to n}^{\rm lq}\end{matrix}.
\end{equation}
\begin{proof} See Appendix A.\end{proof}

\textbf{\textit{Proposition 2:}} Define the queuing delay of all $N$ DNN tasks due to the on-device inference  for the $n$-th DNN task, i.e., $D^{\rm lq}_n=\sum_{m=1}^{N} D_{n\to m}^{\rm lq}$, as the long-term on-device queuing delay of the $n$-th DNN task, which can be calculated as follows:\looseness=-1
\begin{equation}\label{Dn_lq_new_def}
D^{\rm lq}_n=\left\{
\begin{aligned}
&0, ~T_n^{\rm lc}=0,
\\&\begin{matrix}\sum_{t=t_{n,0}}^{t_{n,0}+(T_n^{\rm lc}/\Delta T)-1} Q^{D}(t)\Delta T\end{matrix},~\rm{otherwise}.
\end{aligned}
\right.
\end{equation}

\begin{proof} See Appendix B. \end{proof}

Equation (\ref{Dn_lq_new_def}) can be interpreted as follows. First, if a DNN task is directly offloaded to the edge server without any on-device inference, i.e., $T_n^{\rm lc}=0$, the on-device queuing delay of the other tasks would not be increased, i.e., $D^{\rm lq}_n=0$. Otherwise, if the device is processing the task in the $t$-th time slot, the on-device queuing delay of each task in the on-device task queue will be increased by the time duration of a time slot, i.e., $\Delta T$. As a result, the sum of the increased on-device queuing delay for the tasks in the on-device queue in the $t$-th time slot is $Q^{D}(t)\Delta T$. The increased queuing delay of the subsequent tasks due to the on-device inference for the $n$-th task is thus the summation term in (\ref{Dn_lq_new_def}).

By substituting $T_n^{\rm lq}(x_n)$ in the overall delay (\ref{queuing_delay}) with $D_n^{\rm lq}(x_n)$, we define the time cost of the $n$-th task as:
\begin{equation}
C_n(\mathbf{x}_n) = D^{\rm lq}_n(\mathbf{x}_{n})+T^{\rm lc}_n(x_n) +T_n^{\rm up}(x_n)
+T_n^{\rm eq}(\mathbf{x}_n)+ T^{\rm ec}(x_n),
\end{equation}
where $D^{\rm lq}_n$ is a function of $\mathbf{x}_{n}$ since $t_{n,0}$ and $T_n^{\rm lc}$ in (\ref{Dn_lq_new_def}) are functions of $\mathbf{x}_{n-1}$ and $x_n$, respectively.

By substituting $T_n(\mathbf{x}_n)$ in (\ref{utility}) with $C_n(\mathbf{x}_n)$, we define the long-term utility of the $n$-th DNN task by:
\begin{equation}\label{newutility}
U^{\rm lt}_n(\mathbf{x}_n) = -C_n(\mathbf{x}_n)+\alpha A_n(x_n)-\beta E_n(x_n).
\end{equation}

Compared with the original utility of a DNN task in (\ref{utility}), which incorporates the on-device queuing delay of the task, the long-term utility in (\ref{newutility}) incorporates the on-device queuing delay of subsequent tasks due to the on-device inference for the task. As a result, when sequentially and individually maximizing the long-term utility of each task, the on-device queuing delay of subsequent tasks can be minimized accordingly.

Based on \textit{Proposition 1}, the sum of the on-device queuing delay of all the $N$ DNN tasks can be calculated by:
\begin{equation}\label{sum_D_nlq}
\begin{aligned}
\begin{matrix}\sum_{n=1}^{N}D^{\rm lq}_n\end{matrix}&~=~ \begin{matrix}\sum_{n=1}^{N}(\sum_{m=1}^{N} D_{n\to m}^{\rm lq})\end{matrix}
\\&~=~ \begin{matrix}\sum_{n=1}^{N}(\sum_{m=1}^{N} D_{m\to n}^{\rm lq}\end{matrix})
\\&~=~ \begin{matrix}\sum_{n=1}^{N}T^{\rm lq}_{n}\end{matrix}.
\end{aligned}
\end{equation}
Thus, we have:
\begin{equation}
\begin{matrix}\sum_{n=1}^{N} U_n(\mathbf{x}_n)=\sum_{n=1}^{N} U^{\rm lt}_n(\mathbf{x}_n),\end{matrix}
\end{equation}
which indicates that the average utility of the $N$ DNN tasks, as the objective in ${\rm(P1)}$, equals to the average long-term utility of the $N$ DNN tasks.

\textit{2) Problem Transformation:} 
Due to the stochastic DNN task generation at the considered device and the task arrival at the edge server, we transform ${\rm(P1)}$ into the sequential and individual maximization of the \textit{expected} long-term utility of a task given the offloading decisions of all preceding tasks:
\begin{equation}\label{problem_decompose}
{\rm(P2)}~\bm{\max}_{\widehat{x}_{n}(\mathbf{x}_{n-1})\leq x_n\leq l_e+1}~ \mathbb{E} \left[U^{\rm lt}_n(\mathbf{x}_n)\right].
\end{equation}

\section{DT and Learning-Assisted Algorithm for Offloading decision-making}\label{DT-assited decision_chap}
\begin{figure*}
    \centering
    \includegraphics[width=18cm]{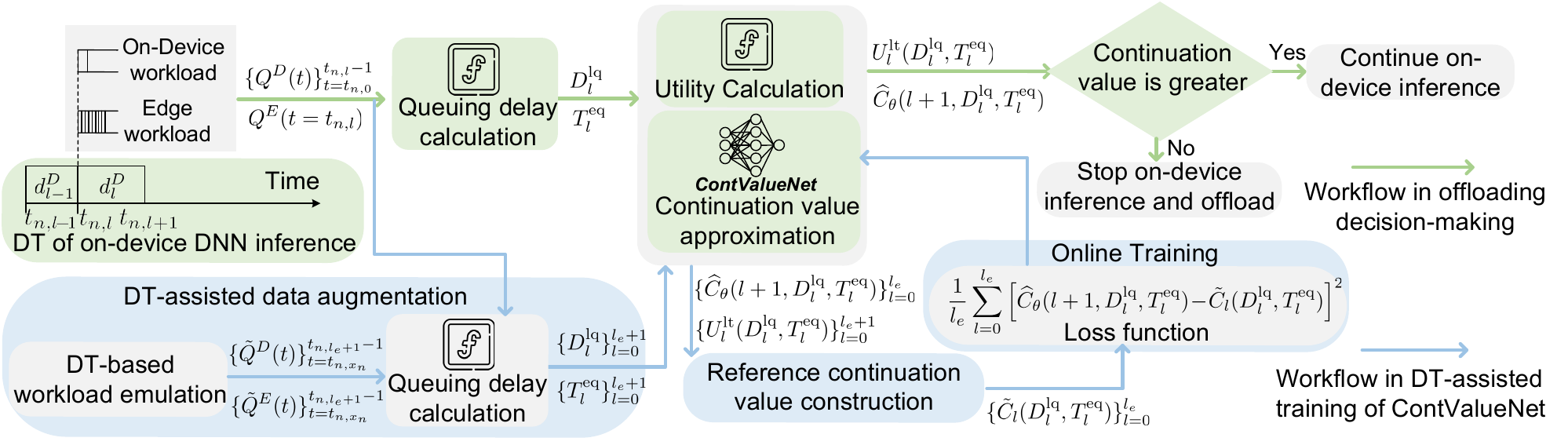}
    \caption{DT and learning-assisted algorithm for DNN task offloading decision-making.}
    \label{strategyoverview}
\end{figure*}

In this section, to solve the transformed problem ${\rm(P2)}$, we propose an algorithm for DNN task offloading decision-making, assisted by DTs and machine learning techniques as shown in Fig. \ref{strategyoverview}. In this algorithm, we use the optimal stopping theorem and a neural network to make offloading decisions, and propose DT-assisted training for the neural network.

For brevity, we denote $D_n^{\rm lq}(x_n\!=\!l|\mathbf{x}_{n-1})$, $T_n^{\rm eq}(x_n\!=\!l|\mathbf{x}_{n-1})$ and $U_n^{\rm lt}(x_n=l|\mathbf{x}_{n-1})$ by $D_{l}^{\rm lq}$, $T_{l}^{\rm eq}$ and $U^{\rm lt}_{l}$, which respectively represent the long-term on-device queuing delay, the edge queuing delay and the long-term task utility if the offloading decision of the $n$-th DNN task is $l$, i.e., $x_n=l$, given that the offloading decisions of the first $n\!-\!1$ DNN tasks are $\mathbf{x}_{n-1}$. To emphasize the impact of the long-term on-device queuing delay and the edge queuing delay on the long-term utility of a DNN task, we represent $U^{\rm lt}_{l}$ as $U^{\rm lt}_{l}(D_{l}^{\rm lq},T_{l}^{\rm eq})$. In addition, we define $\mathbf{D}_l^{\rm lq} =\{D_{0}^{\rm lq},D_{1}^{\rm lq},..., D_{l}^{\rm lq}\}$ and $\mathbf{T}_l^{\rm eq} =\{T^{\rm eq}_{0},T^{\rm eq}_{1},...,T^{\rm eq}_{l}\}$.

\subsection{DT and Learning-Assisted Offloading Decision-Making}\label{DT_learning_offloading}

\textit{1) Continuation Value-based Offloading Decision-Making:} Define $V_{l}(\mathbf{D}_{l}^{\rm lq}, \mathbf{T}_{l}^{\rm eq})$ as the \textit{maximum} expected long-term utility of a DNN task when $l$ layers of the shallow DNN have been executed for the task, where $l\!\in\{0,1,...,l_e+1\}$. Such a value satisfies a recursion rule:
\begin{equation}\label{value_f}
V_{l}(\mathbf{D}_{l}^{\rm lq},\mathbf{T}_{l}^{\rm eq})\!=\!
\left\{
\begin{aligned}
&\max\Big\{U^{\rm lt}_{l}(D_{l}^{\rm lq}, T_{l}^{\rm eq}), C_{l}(\mathbf{D}_{l}^{\rm lq}, \mathbf{T}_{l}^{\rm eq})\Big\},l\leq l_e,
\\&U^{\rm lt}_{l_e+1}(D_{l_e+1}^{\rm lq}, T_{l_e+1}^{\rm eq}), l=l_e+1,
\end{aligned}
\right.
\end{equation}
where $\forall l\in\{0,1,...,l_e\}$, we have
\begin{equation}\label{contValue_rule}
C_{l}(\mathbf{D}_{l}^{\rm lq}, \mathbf{T}_{l}^{\rm eq}) = \mathbb{E}\Big[V_{l+1}(\mathbf{D}_{l+1}^{\rm lq}, \mathbf{T}_{l+1}^{\rm eq})|\mathbf{D}_l^{\rm lq},\mathbf{T}_l^{\rm eq}\Big],
\end{equation}
referred to as the \textit{continuation value}\cite{herrera2021optimal}. 

\textbf{\textit{Proposition 3:}} Based on optimal stopping \cite{thomasOptimalStoppingApplications}, the optimal offloading decision for maximizing the expected long-term utility of the $n$-th DNN task is:
\begin{subequations}\label{opt_stopping}
\begin{equation}
x_n
\!=\!\left\{
\begin{aligned}
&\min\{\Psi\!\}, ~{\rm if}~\Psi^{\rm} \neq \emptyset,
\\&l_e+1, ~{\rm otherwise},
\end{aligned}
\right.
\end{equation}%
where
\begin{equation}\label{opt_stopping_contValue}
\Psi = \Big\{l\mid\widehat{x}_n \!\leq \!l\! \leq l_e,U^{\rm lt}_{l}(D_{l}^{\rm lq}, T_{l}^{\rm eq}) \!\geqslant\! C_{l}(\mathbf{D}_{l}^{\rm lq}, \mathbf{T}_{l}^{\rm eq})\Big\}.
\end{equation}
\end{subequations}
The continuation value-based offloading decision-making given in (\ref{opt_stopping}) is explained as follows. For the $n$-th DNN task, after the device executes $l$ layers of the shallow DNN, where $l \in \{\widehat{x}_n,\widehat{x}_n+1,..., l_e\}$, if the long-term utility of offloading the DNN task now, i.e., $U^{\rm lt}_{l}(D_{l}^{\rm lq}, T_{l}^{\rm eq})$, is no less than the continuation value, i.e., $C_{l}(\mathbf{D}_{l}^{\rm lq}, \mathbf{T}_{l}^{\rm eq})$, the on-device inference for the task should be stopped for offloading the task to the edge server. Otherwise, the device should continue executing the ($l+1$)-th layer of the shallow DNN.


\textit{2) DT and Learning-Assisted Offloading Decision-Making:} The continuation value for each layer of the shallow DNN can be obtained using backward induction \cite{thomasOptimalStoppingApplications}, which however requires the prior statistics of the workload evolution and introduces computing overhead \cite{becker2019deep}. Inspired by \cite{herrera2021optimal}, our approach is to construct a neural network referred to as ContValueNet to generate an approximated continuation value $\widehat{C}_{\theta}(l+1,D_{l}^{\rm lq},T_{l}^{\rm eq})$ given the input $\{l+1,D_{l}^{\rm lq},T_{l}^{\rm eq}\}$, where $\theta$ represents the parameters in ContValueNet. By substituting the continuation value in (\ref{opt_stopping_contValue}) with the approximated continuation value, a learning-assisted offloading algorithm is obtained. Specifically, as shown in the green blocks in Fig. \ref{strategyoverview}, through the DT of on-device DNN inference given in (\ref{exeuction_time_est}), the controller is informed when a layer of the shallow DNN is about to be executed at the device. Before the on-device execution of the ($l+1$)-th layer by the device, the controller calculates \textit{i)} the queuing delay $D_{l}^{\rm lq}$ and $T_{l}^{\rm eq}$; \textit{ii)} calculates the long-term utility of the task $U^{\rm lt}_{l}(D_{l}^{\rm lq}, T_{l}^{\rm eq})$; and \textit{iii)} approximates the continuation value by the ContValueNet, i.e., $\widehat{C}_{\theta}(l+1,D_{l}^{\rm lq},T_{l}^{\rm eq})$. If the continuation value is greater than the utility, the controller lets the device execute the next layer for the task. Otherwise, the network controller lets the device upload the input data of the ($l+1$)-th layer in the full-size DNN to the edge server.\looseness=-1

\subsection{DT-Assisted Training of ContValueNet}\label{DT-augment}
As shown in the blue blocks in Fig. \ref{strategyoverview}, DTs are leveraged to assist in training ContValueNet, which involves three steps, i.e., DT-assisted data augmentation, reference continuation value construction, and online ContValueNet training.



\textit{1) DT-Assisted Data Augmentation:} If the $n$-th DNN task is determined to offload to the edge server, i.e., $x_n \leq l_e$, the controller emulates the long-term on-device queuing delay and the edge queuing delay given the remaining potential offloading decisions for the task, i.e., $\{{D}_{l}^{\rm lq}\}_{l=x_n+1}^{l_e+1}$ and $\{{T}_{l}^{\rm eq}\}_{l=x_n+1}^{l_e+1}$, using the DT of workload evolution and (\ref{dt_emulation}). Specifically, $\{{D}_{l}^{\rm lq}\}_{l=x_n+1}^{l_e}$ are calculated by substituting $Q^{D}(t)$ in (\ref{Dn_lq_new_def}) with $\tilde{Q}^{D}(t)$ in (\ref{emu_devicequeue}), and $\{T_{l}^{\rm eq}\}_{l=x_n+1}^{l_e+1}$ are calculated by substituting the on-device workload in (\ref{edge_queuing}) with the emulated on-device workload in (\ref{emu_edgequeue}). Then, the controller calculates the approximated continuation values $\{\widehat{C}_{\theta}(l+1,D_{l}^{\rm lq},T_{l}^{\rm eq})\}_{l=x_n+1}^{l_e+1}$ using ContValueNet. Combining the above values with $\{{D}_{l}^{\rm lq}\}_{l=0}^{x_n}$, $\{{T}_{l}^{\rm eq}\}_{l=0}^{x_n}$ and $\{\widehat{C}_{\theta}(l+1,D_{l}^{\rm lq},T_{l}^{\rm eq})\}_{l=0}^{x_n}$ which are collected during the decision-making for the $n$-th DNN task, the controller obtains $\{{D}_{l}^{\rm lq}\}_{l=0}^{l_e+1}$, $\{{T}_{l}^{\rm eq}\}_{l=0}^{l_e+1}$ and $\{\widehat{C}_{\theta}(l+1,D_{l}^{\rm lq},T_{l}^{\rm eq})\}_{l=0}^{l_e+1}$, which will be used to construct the reference continuation values.

\textit{2) Reference Continuation Value Construction:} To optimize the parameters $\theta$ of ContValueNet, reference continuation values of evaluating the approximation error of ContValueNet should be constructed \cite{herrera2021optimal}.

First, for any $l \in \{0,1,2,...,l_e\}$, we calculate $\widehat{V}_{l}(D_{l}^{\rm lq},T_{l}^{\rm eq})$ by substituting the continuation value in (\ref{value_f}) with the approximated continuation value:
\begin{equation}\label{value_f_appro}
\widehat{V}_l(D_{l}^{\rm lq},T_{l}^{\rm eq})=\max\Big\{U^{\rm lt}_{l}(D_{l}^{\rm lq}, T_{l}^{\rm eq}),
\widehat{C}_{\theta}(l+1,D_{l}^{\rm lq},T_{l}^{\rm eq})\Big\}.
\end{equation}

Then, the reference continuation value can be obtained according to (\ref{contValue_rule}):
\begin{equation}\label{continue_approx}
\tilde{C}_l(D_{l}^{\rm lq},T_{l}^{\rm eq})= \mathbb{E}\Big[\widehat{V}_{l+1}(D_{l+1}^{\rm lq}, T_{l+1}^{\rm eq})|\mathbf{D}_l^{\rm lq},\mathbf{T}_l^{\rm eq}\Big].
\end{equation}
Due to the unknown statistics, we use the single-sample estimation method \cite{935083} to approximate the expectation term in (\ref{continue_approx}):
\begin{equation}\label{expect_approx}
 \mathbb{E}\Big[\widehat{V}_{l+1}(D_{l+1}^{\rm lq}, T_{l+1}^{\rm eq})|\mathbf{D}_l^{\rm lq},\mathbf{T}_l^{\rm eq}\Big]\approx \widehat{V}_{l+1}(D_{l+1}^{\rm lq},T_{l+1}^{\rm eq}).
\end{equation}

Finally, by combining (\ref{continue_approx}) and (\ref{expect_approx}), and calculating $\widehat{V}_{l+1}(D_{l+1}^{\rm lq},T_{l+1}^{\rm eq})$ in (\ref{expect_approx}) according to (\ref{value_f_appro}), the reference continuation value is obtained by:
\begin{equation}\label{continue_ref}
\tilde{C}_l(D_{l}^{\rm lq},T_{l}^{\rm eq})\!=\! \max\Big\{U^{\rm lt}_{l+1}(D_{l+1}^{\rm lq}, T_{l+1}^{\rm eq}),
\!\widehat{C}_{\theta}(l+2,D_{l+1}^{\rm lq},T_{l+1}^{\rm eq})\Big\}\!.
\end{equation}

\textbf{\textit{Remark 1:}} To calculate the reference continuation value $\tilde{C}_l(D_{l}^{\rm lq},T_{l}^{\rm eq})$ in (\ref{continue_ref}), $D_{l+1}^{\rm lq}$ and $T_{l+1}^{\rm eq}$ are required. As a result, with $\{{D}_{l}^{\rm lq}\}_{l=0}^{x_n}$ and $\{{T}_{l}^{\rm eq}\}_{l=0}^{x_n}$ collected in the offloading decision-making for the $n$-th task, reference continuation values before executing each of the first $x_n$ layers of the shallow DNN can be obtained, where $x_n \!\leq \!l_e+1$. By contrast, with DT-assisted data augmentation, which generates $\{{D}_{l}^{\rm lq}\}_{l=x_n+1}^{l_e+1}$ and $\{{T}_{l}^{\rm eq}\}_{l=x_n+1}^{l_e+1}$,  reference continuation values before executing each of the $l_e+1$ layers in the shallow DNN can be obtained. In this way, the data for training ContValueNet is augmented. 

\textit{3) Online ContValueNet Training:} 
To mitigate the impact of the one-sample estimation method in (\ref{expect_approx}), ContValueNet is trained in an online manner with a carefully designed loss function. Specifically, the loss function for training ContValueNet after processing the first $\widehat{n}$ DNN tasks, where $1\leq \widehat{n}\leq M$, is denoted by $L^{\rm NN} (\theta,\widehat{n})$ and defined by:
\begin{equation}\label{loss_fun}
\begin{split}
&L^{\rm NN} (\theta,\widehat{n})
\\&= \frac{1}{\widehat{n}(l_e+1)}\sum_{n=0}^{\widehat{n}}\!\sum_{l=0}^{l_e}\Big[\widehat{C}_{\theta}(l+1,D_{l}^{\rm lq},T_{l}^{\rm eq})\!-\!\tilde{C}_l(D_{l}^{\rm lq},T_{l}^{\rm eq})\Big]^2,
\end{split}
\end{equation}
which is the mean squared error of the continuation value approximation.
Then, the gradient descent method is applied to optimize the parameters $\theta$ in ContValueNet to minimize the loss function. Specifically, the update rule of $\theta$ is:
\begin{equation}
\theta^{\prime} = \theta- \gamma \nabla_{\theta} L^{\rm NN} (\theta,\widehat{n}),
\end{equation}
where $\gamma$ is the learning rate; $\nabla_{\theta} L^{\rm NN} (\theta,\widehat{n})$ is the first-order derivative of the loss function with respect to $\theta$.


\section{Offloading Decision Space Reduction}\label{decison_reduction_chap}

The complexity of the approach to adaptive device-edge collaboration on DNN inference in Section \ref{DT_learning_offloading} can be high due to the potentially large number of layers in the DNN. 
In this section, from two aspects, we will derive necessary conditions for an offloading decision to be optimal in maximizing the expected long-term utility. Correspondingly, we investigate decision space reduction to reduce the complexity. 

\subsection{DNN Layers with Negligible Execution Time}
The execution time of some layers in a DNN, e.g., pooling layers in a convolutional neural network (CNN), is negligible due to relatively simple operations \cite{pooling_negligible_trans}. As a result, such a layer has a negligible impact on the long-term on-device queuing delay and the edge queuing delay of a DNN task. By contrast, the data size before and after the execution of such layers can be drastically changed. For example, a max-pooling layer downsamples input data to learn spatially robust features for image recognition tasks, while a max-unpooling layer upsamples the input data to restore the original data size for semantic segmentation tasks.


\textbf{\textit{Remark 2:}} If a layer in a DNN has negligible execution time and outputs data with reduced (increased) size, offloading the DNN task before (after) the on-device execution of the layer cannot be optimal for maximizing the expected long-term utility of the DNN task. This is because offloading a DNN task after the data size is increased or before the data size is reduced would result in a higher data uploading delay. Based on this necessary condition, we can treat each max-pooling layer and its preceding layer as one logical layer and each max-unpooling layer and its succeeding layer as one logical layer in the offloading decision-making for a DNN task. 



\subsection{Properties of Workload Evolution}
Based on the properties in the evolution of the on-device workload and the edge server workload over time, properties in the evolution of long-term on-device queuing delay and edge queuing delay (as the on-device inference continues) can be derived, respectively.

\textbf{\textit{Property 1:}} During the on-device inference for a DNN task, the on-device workload $Q^{D}(t)$ is non-decreasing with $t$. In addition, the on-device inference delay $T_n^{\rm lc}$ increases with $x_n$. As a result, when $x_n$ increases, the long-term on-device queuing delay $D^{\rm lq}_n(x_n)$, calculated by (\ref{Dn_lq_new_def}), is non-decreasing with $x_n$. The increase of $D^{\rm lq}_n(x_n)$ is minimized if no DNN tasks are generated after the controller starts to make the offloading decision for the $n$-th DNN task, i.e., $Q^{D}(t)=Q^{D}(t_{n,\widehat{x}_n}), \forall t\geq t_{n,\widehat{x}_n}$. 
The minimum increase is equal to the product of $Q^{D}(t_{n,\widehat{x}_n})$ and the increase of on-device inference delay $T^{\rm lc}_n(x_n)$.

\textbf{\textit{Property 2:}} The edge server workload can decrease or increase during the extended on-device inference for the $n$-th task due to the increase of $x_n$. The decrease of the workload is maximized if no tasks arrive at
 the edge server during the extended on-device inference, and the maximum decrease is equal to the workload that the edge server can process during the extended on-device inference. As a result, given the increase of $x_n$, the maximum decrease of the edge queuing delay $T^{\rm eq}_n(x_n)$, calculated by dividing the edge server workload by the edge server processing capability, equals to the increase of the on-device inference delay $T^{\rm lc}_n(x_n)$.

Using the above properties, a necessary condition for an offloading decision $x_n \in \{\widehat{x}_n,...,l_e\}$ to be optimal can be derived. 

\textbf{\textit{Lemma 1:}} Define a deterministic part in the long-term utility of the $n$-th DNN task as $U_n^{\rm pt}(x_n) \overset{def}{=} -T_n^{\rm up}(x_n)\!-\!T_n^{\rm ec}(x_n)-\!\beta E_n(x_n)$. If $x_n^{*}\le l_e$ maximizes $\mathbb{E} \left[U_n^{\rm lt}(x_n)\right]$, then for any $x_n\in \{\widehat{x}_n,..., x_n^*\}$, it holds:
\begin{equation}\label{lemma6.2}
U_n^{\rm pt}(x_n^*)\ge U_n^{\rm pt}(x_n)+Q^{D}(t\!=\!t_{n,\widehat{x}_n})(T^{\rm lc}_n(x_n^*)-T^{\rm lc}_n(x_n)),
\end{equation}
where $t_{n,\widehat{x}_n}$ is the index of the time slot from which the $n$-th DNN task can be offloaded to the edge server.
\begin{proof}Note that $U_n^{\rm lt}(x_n) = U_n^{\rm pt}(x_n)-T^{\rm lc}_n(x_n)- D^{\rm lq}_n(x_n) - T^{\rm eq}_n(x_n)+\alpha A_n(x_n)$. If $x_n^{*}\le l_e$ maximizes $\mathbb{E} \left[U_n^{\rm lt}(x_n)\right]$, then for any $x_n\in \{\widehat{x}_n,..., x_n^*\}$, we have:
\begin{equation}\label{step1}
\begin{split}
&U_n^{\rm pt}(x_n^*)\!-\!T^{\rm lc}_n(x_n^*)\!-\!\mathbb{E}\left[D^{\rm lq}_n(x_n^*)\right]\! -\! \mathbb{E}\left[T^{\rm eq}_n(x_n^*)\right]\!+\!\alpha A_n(x_n^*)
\\&\ge U_n^{\rm pt}(x_n)\!-\!T^{\rm lc}_n(x_n)\!-\!\mathbb{E}\left[D^{\rm lq}_n(x_n)\right]\!-\! \mathbb{E}\left[T^{\rm eq}_n(x_n)\right]\!+\!\alpha A_n(x_n).
\end{split}
\end{equation}

When $x_n\!\le\! l_e$ (including the case when $x_n^*\!\le\! l_e$), the $n$-th DNN task is offloaded to the edge server for processing by the full-size DNN. The inference accuracy is thus the same under the two cases:
\begin{equation}\label{step4}
A_n(x_n^*)= A_n(x_n).
\end{equation}

According to \textit{Property 1}, we have:
\begin{equation}\label{step2}
D^{\rm lq}_n(x_n^*)\ge D^{\rm lq}_n(x_n)+Q^{D}(t=t_{n,\widehat{x}_n})(T^{\rm lc}_n(x_n^*)-T^{\rm lc}_n(x_n)).
\end{equation}

According to \textit{Property 2}, we have:
\begin{equation}\label{step3}
T_{n}^{\rm lc}(x_n^*)-T_{n}^{\rm lc}(x_n)\ge T^{\rm eq}_n(x_n)-T^{\rm eq}_n(x_n^*).
\end{equation}

Combining (\ref{step1}), (\ref{step4}), (\ref{step2}) and (\ref{step3}), (\ref{lemma6.2}) is obtained.
\end{proof}


In addition, we derive a necessary condition of the device-only inference, i.e., $x_n=l_e+1$, to be optimal in maximizing the \textit{long-term utility} $U_n^{\rm lt}(x_n)$.\footnote{The long-term utility of a task by device-only inference would not be affected by the dynamic workload at the edge server, while the utility by the other inference scenarios would be affected. Due to the unknown statistics of the dynamic workload, we cannot derive any necessary condition for device-only inference to be optimal in maximizing the \textit{expected} long-term utility.\looseness=-1}


\textbf{\textit{Lemma 2:}}  If $x_n=l_e+1$ maximizes the long-term task utility $U_n^{\rm lt}(x_n)$, then:
\begin{equation}\label{lemma1.3}
U_n(l_e+1)\ge U_n(\widehat{x}_n)+Q^{D}(t\!=\!t_{n,\widehat{x}_n})(T^{\rm lc}_n(l_e+1)-T^{\rm lc}_n(\widehat{x}_n)).
\end{equation}
\begin{proof}
Since $x_n=l_e+1$ maximizes the long-term task utility $U_n^{\rm lt}(x_n)$, we have:
\begin{equation}\label{equ_on_device_opt}
U_n^{\rm lt}(l_e+1)\ge U_n^{\rm lt}(\widehat{x}_n).
\end{equation}
According to \textit{Property 1}, we have:
\begin{equation}\label{min_increase_lt_queuing}
D^{\rm lq}_n(l_e+1)\!\ge\! D^{\rm lq}_n(\widehat{x}_n)+Q^{D}(t=t_{n,\widehat{x}_n})(T^{\rm lc}_n(l_e+1)-T^{\rm lc}_n(\widehat{x}_n)).
\end{equation}
Note that:
\begin{subequations}
\begin{equation}
U_n^{\rm lt}(l_e+1)+D^{\rm lq}_n(l_e+1)=U_n(l_e+1)+T_n^{\rm lq}(\mathbf{x}_{n-1}),
\end{equation}
\begin{equation}
U_n^{\rm lt}(\widehat{x}_n)+D^{\rm lq}_n(\widehat{x}_n)=U_n(\widehat{x}_n)+T_n^{\rm lq}(\mathbf{x}_{n-1}).
\end{equation}
\end{subequations}
Combining (\ref{equ_on_device_opt}) and (\ref{min_increase_lt_queuing}), (\ref{lemma1.3}) is obtained.
\end{proof}

Based on \textit{Lemma 1} and \textit{Lemma 2}, an algorithm for reducing the offloading decision space is proposed in Algorithm \ref{alg_1}. Specifically, when the $n$-th DNN task can be offloaded to the edge server, the potential offloading decisions $x_n \in \{\widehat{x}_n,\widehat{x}_n+1,...,l_e\}$ will be checked, and only the decisions satisfying the condition given in \textit{Lemma 1} will be considered in the learning-assisted offloading decision-making. In addition, if all the possible offloading decisions $x_n \in \{\widehat{x}_n+1,...,l_e\}$ violate the necessary condition given in \textit{Lemma 1}, the condition in \textit{Lemma 2} will be checked to determine whether device-only inference, i.e., $x_n=l_e+1$, can be optimal for maximizing the long-term task utility. If not, $x_n=l_e+1$ will not be considered. Finally, only the remaining decisions, i.e, $\mathcal{L}_n$, will be considered.

\begin{algorithm}[htbp]
    \DontPrintSemicolon
    \textbf{Input}: $\widehat{x}_n$;\\
    \textbf{Output}: $\mathcal{L}_n$;\\ 
    \textbf{Initialization}: $\mathcal{L}_n=\{\widehat{x}_n,\widehat{x}_n+1,...,l_e+1\}$;\\
    \For{$l \in \{\widehat{x}_n,\widehat{x}_n+1,...,l_e\}$}{
        ~Given $x_n^*=l$, for any $x_n\in \{\widehat{x}_n,..., x_n^*\}$, if (\ref{lemma6.2}) is violated, delete $l$ in the set $\mathcal{L}_n$;
    }
    \If {$\mathcal{L}_n =\{\widehat{x}_n,l_e+1\}$}
        {\If {\rm {(\ref{lemma1.3}) is violated}}{
    delete $l_e+1$ in the set $\mathcal{L}_n$;}
    }
    \Return $\mathcal{L}_n$
    \caption{Decision Space Reduction}
    \label{alg_1}
\end{algorithm}

\section{Simulation Results}\label{simulation_chap}
In this section, through simulations, we evaluate the performance of the DT-assisted approach to adaptive device-edge collaboration on DNN inference.
\subsection{Simulation Settings and Benchmarks}
\begin{figure}
    \centering
    \includegraphics[width=8.5cm]{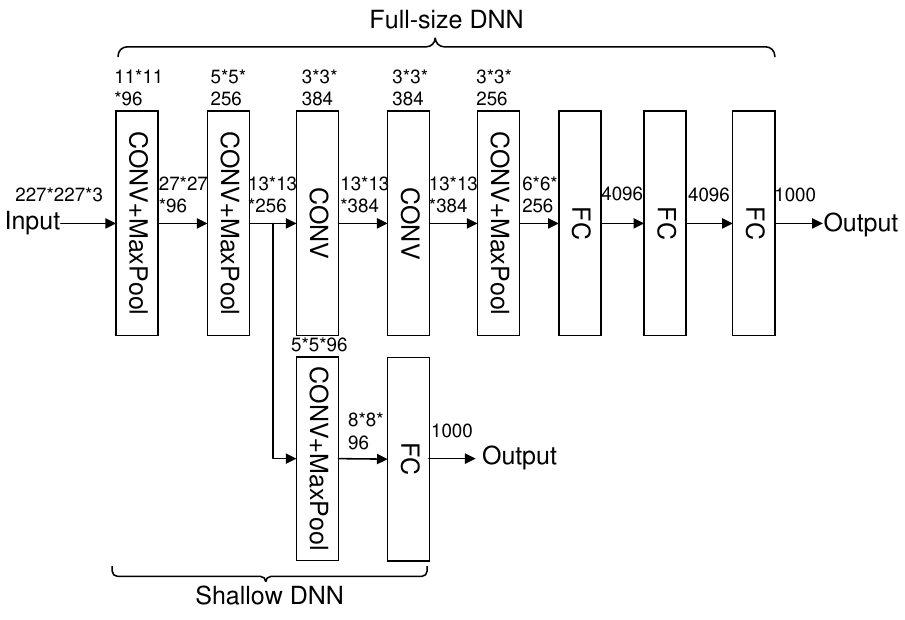}
    \caption{Full-size DNN and shallow DNN.}
    \label{alexnet}
\end{figure}

The main parameters used in the simulations are given in Table \ref{table_1}, and the detailed configurations of the full-size and shallow DNNs are shown in Fig. \ref{alexnet}, where Alexnet is chosen as the full-size DNN. Based on \textit{Remark 2}, we consider each pooling layer and its preceding layer as one logical layer. The execution time for the layers of the shallow DNN by the device, i.e., $\{d_l^D\}_{l=1}^{l_e+1}$, and that for the layers of the full-size DNN by the edge server, i.e., $\{d_l^E\}_{l=1}^{L}$, are estimated based on the number of FLOPs for the layers and the computation frequency of the AIoT device and the edge server\cite{FLops_calculation_JSAC}. 

\begin{table}[htbp]
\caption{Simulation Parameters}
\centering
\begin{tabular}{|l|c|c|}
\hline
  \bf{Parameters} &  \bf{Symbol} &  \bf{Value}\\ \hline
  Time duration of a time slot &  $\Delta T$ & 10 ms\\ \hline
   Exit layer index & $l_e$ &  2    \\      \hline
   Computation frequency of the edge server & $f^E$ & 50 GHz\\ \hline
   Computation frequency of the AIoT device& $f^D$ & 1 GHz\\ \hline
   Accuracy of full-size DNN& $\eta^E$ & 0.9\\ \hline
   Accuracy of shallow DNN& $\eta^D$ & 0.6\\ \hline
   Uplink transmission rate & $R_0$ & 126 Mbps\\ \hline
   Transmit power of the AIoT device& $p^{\rm up}$ & 20 dBm\\ \hline
   Energy coefficient &$\kappa^E, \kappa^D$ & $10^{-30}$\\ \hline
   Weight for accuracy & $\alpha$&1.0\\ \hline
   Weight for energy consumption& $\beta$&0.2\\ \hline
\end{tabular}
\label{table_1}
\end{table}

The neural network for approximating the continuation values, i.e., ContValueNet, consists of three fully-connected layers with $200$, $100$ and $20$ neurons, respectively. For optimizing the parameters therein, the learning rate $\gamma$ is set to be $1\times10^{-3}$ and the Adam optimizer is chosen. In the processing of the first $2000$ DNN tasks, ContValueNet for continuation value approximation is trained, i.e., $M=2000$. The average DNN task utility, inference delay, accuracy and energy consumption are derived by applying the trained ContValueNet to assist making offloading decisions of $8000$ DNN tasks. The DNN task generation at the considered device follows a Bernoulli distribution with probability $p$ and task arrival from the other devices to the edge server follows a Poisson distribution with arrival rate $\lambda$. The CPU cycles required for processing a computing task follows uniform distribution $U (0, U^{\rm max})$, with the value of $U^{\rm max}$ set as $8\times 10^9$. The DNN task generation rate in the unit of tasks per second can be calculated by $p/\Delta$. In addition, the edge processing load can be calculated by $\lambda U^{\rm max}/2f^E$. As a unitless ratio, the edge processing load is not tied to a specific system parameter such as computing capability, task arrival rate, and task computation cost. As a result, using this metric allows our simulation results to offer findings that are applicable across various AIoT network settings.

In the simulations, the following three benchmarks are used: 
\begin{itemize}
\item \textbf{One-Time Ideal Case:} The offloading decision of a DNN task is made only once (upon task generation) to maximize the long-term utility of the task, assuming perfect knowledge of the future on-device workload and edge server workload.
\item \textbf{One-Time Long-term:} The offloading decision of a DNN task is made only once (upon task generation) to maximize the long-term utility of the DNN task based on the current on-device workload and edge server workload.
\item \textbf{One-Time Greedy:} Similar to \cite{kang2017neurosurgeon}, the offloading decision of a DNN task is made only once (upon task generation) to maximize the utility of the DNN task based on the current on-device workload and edge server workload.
\end{itemize}
In addition, we will compare the performance of the learning-assisted task offloading algorithm \textit{i)} with and without the DT-assisted training data augmentation; \textit{ii)} with and without the offloading decision space reduction, respectively.

\begin{figure}
    \centering
    \includegraphics[width=5.5cm]{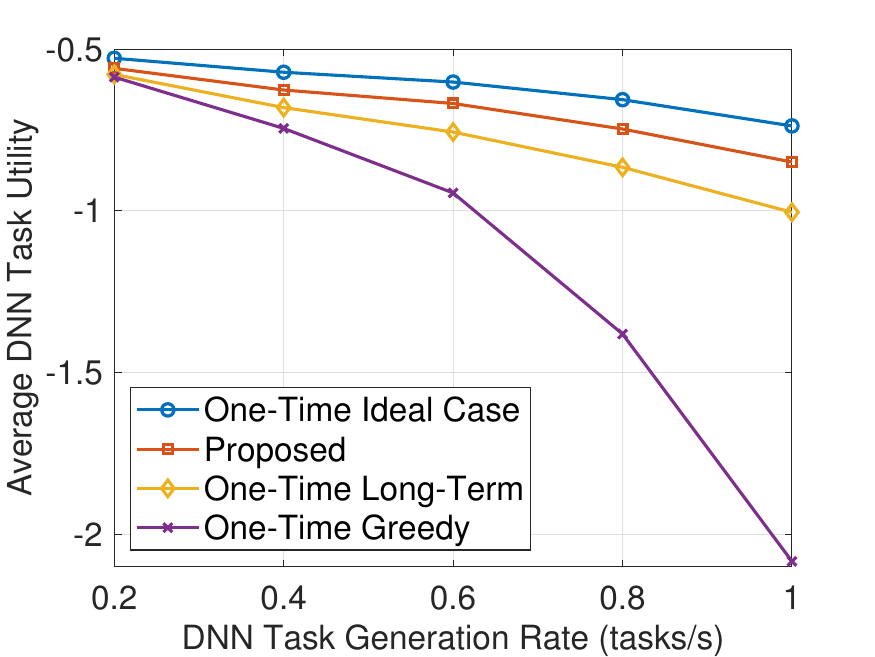}
    \caption{Average DNN task utility versus task generation rate.}
    \label{sim_fig1_ava_utility_versus_DNNrate}
\end{figure}
\begin{figure}
    \centering
    \includegraphics[width=5.5cm]{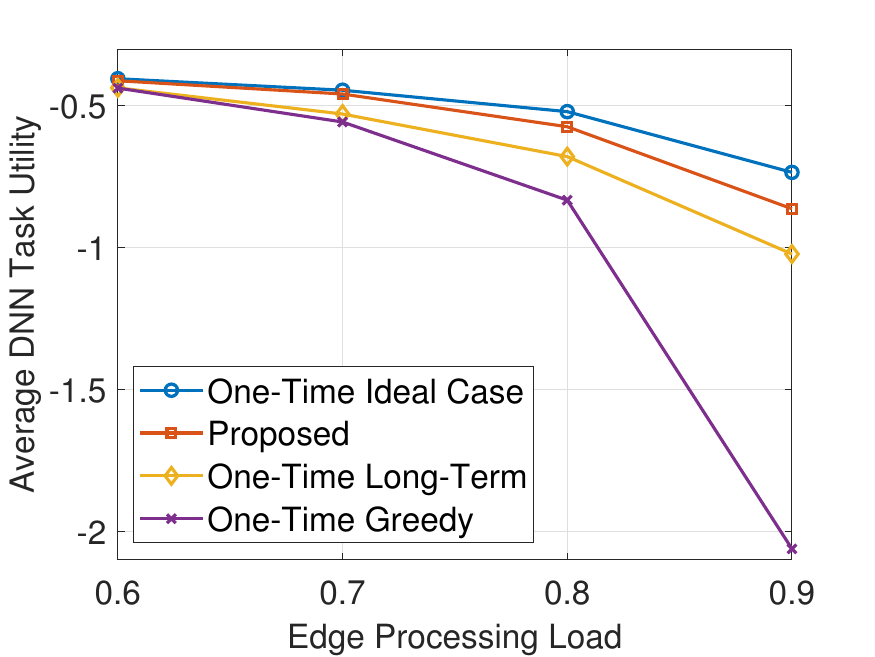}
    \caption{Average DNN task utility versus edge processing load.}
    \label{sim_fig2_ava_utility_versu_edge_utilization}
\end{figure}

\subsection{Adaptiveness to Dynamic Computing Workload}

In Fig. \ref{sim_fig1_ava_utility_versus_DNNrate}, the average utility of DNN tasks under varying DNN task generation rate is shown, where the edge processing load is $0.9$. It can be observed that the proposed approach outperforms the one-time greedy benchmark. This is because the proposed approach makes the offloading decision of a DNN task while considering the on-device queuing delay of subsequent DNN tasks. In addition, the proposed approach outperforms the one-time long-term benchmark, since instead of making task offloading decision upon the task generation, our approach continuously monitors the on-device workload during the on-device inference for each task and adaptively makes the offloading decision.

In Fig. \ref{sim_fig1_ava_utility_versus_DNNrate}, it can be observed that the performance gain compared to the one-time long-term benchmark increases with DNN task generation rate. This is because the increase of the task generation rate results from the increase of the task generation probability, which in turn increases the variance of the Bernoulli distribution and the dynamics of the on-device workload. Since the performance gain comes from better adaptiveness to the dynamics, it increases with the dynamics. In Fig. \ref{sim_fig2_ava_utility_versu_edge_utilization}, the average DNN task utility under varying edge processing load is shown, where the DNN task generation rate is $1.0$. Similarly, the proposed approach outperforms the one-time long-term and one-time greedy benchmarks and the performance gain increases with the edge processing load. 

\begin{figure}
    \centering
    \subfloat[]{\includegraphics[width=5.5cm]{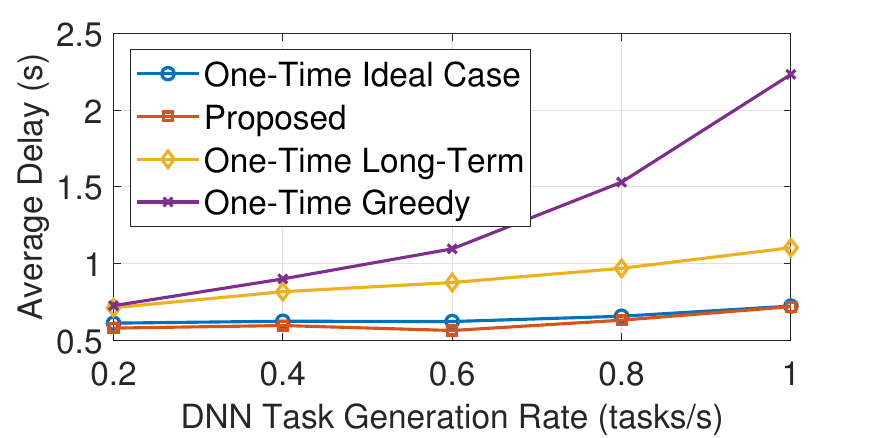}}\\
    \vspace*{-2pt}
    \subfloat[]{\includegraphics[width=5.5cm]{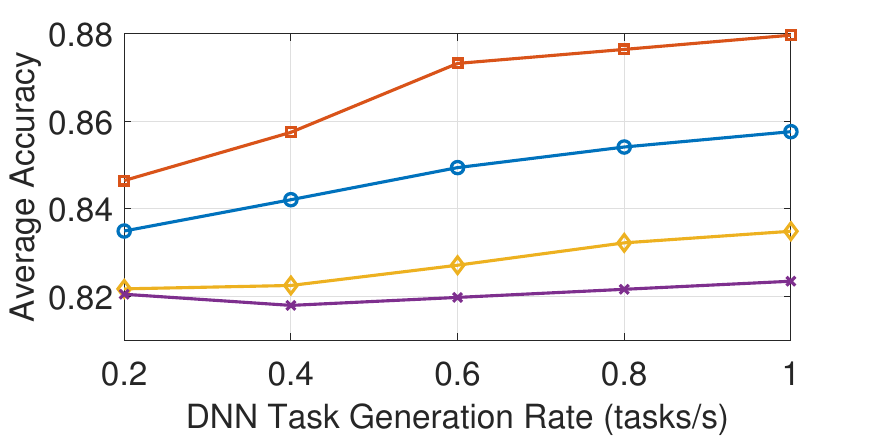}}\\
    \vspace*{-2pt}
    \subfloat[]{\includegraphics[width=5.5cm]{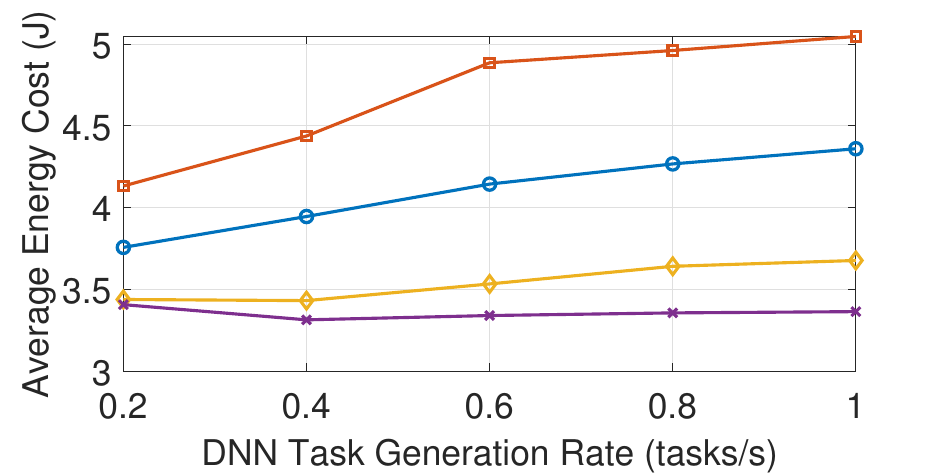}}
    \caption{(a) Average delay; (b) Average accuracy; (c) Average energy consumption versus DNN task generation rate.}
    \label{eachmetric}
\end{figure}

In Fig. \ref{eachmetric}, the average inference delay, accuracy and energy consumption under varying DNN task generation rate is shown, where the edge processing load is $0.9$. Compared with the benchmarks, the proposed approach achieves lower delay and higher inference accuracy at the cost of higher energy consumption. This results from the weights for the delay, inference accuracy and the energy consumption in the task utility, which are set as $1.0, 1.0$, and $0.002$, respectively. Combining the ranges of the three metrics shown in Fig. \ref{eachmetric}, it can be seen that the delay and the accuracy contribute significantly more than the energy consumption in determining the task utility. Since the proposed approach does not assume perfect knowledge of the workload evolution, it can conservatively offload a task to the edge server for reducing the on-device queuing delay of the tasks in the on-device queue and increasing the inference accuracy.



\subsection{Effectiveness of DT-Assisted Training Data Augmentation}

\begin{figure}
    \centering
    \subfloat[]{\includegraphics[width=5.5cm]{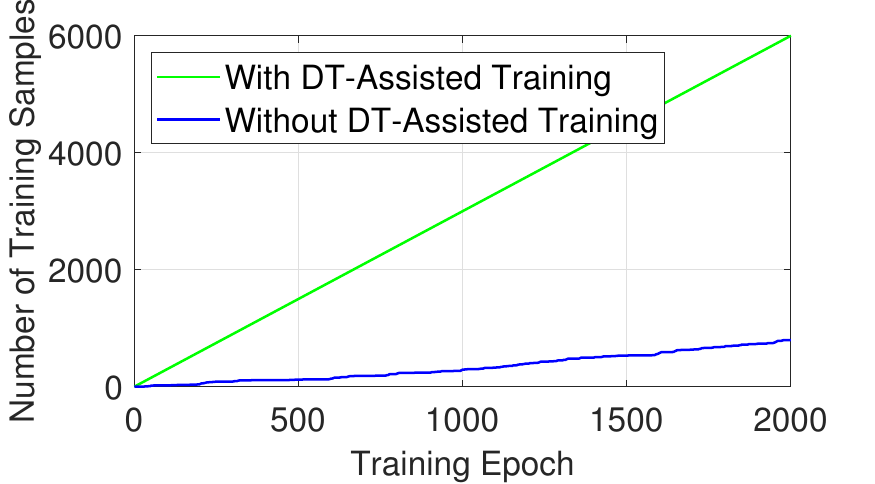}
    \label{sim_fig6_training_samples_versus_epoch_04}}\\
    \vspace*{-2pt}
    \subfloat[]{\includegraphics[width=5.5cm]{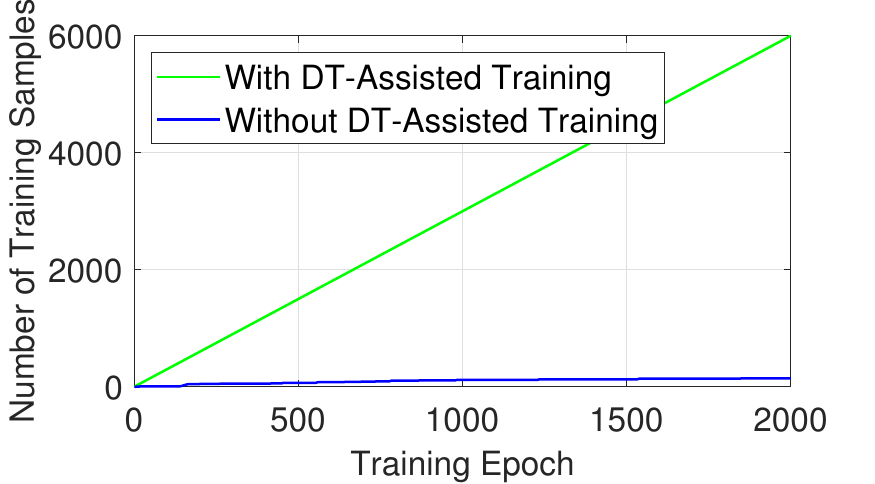}    \label{sim_fig6_training_samples_versus_epoch_08}}
    \vspace*{-2pt}
    \caption{Number of collected training samples with and without DT-assisted training data augmentation.}
    \label{sim_fig6_training_samples_versus_epoch}
\end{figure}

In Fig. \ref{sim_fig6_training_samples_versus_epoch}, the number of training samples collected during the training of ContValueNet is shown, where the edge processing load is $0.9$, and the DNN task generation rate is $0.4$ and $0.8$ in Fig. \ref{sim_fig6_training_samples_versus_epoch_04} and Fig. \ref{sim_fig6_training_samples_versus_epoch_08}, respectively. It can be observed that with DT-assisted training data augmentation, the number of training samples linearly grows as the training continues. This is because the DT of workload evolution is used by the network controller to emulate the on-device and edge server workload under every possible offloading scenario so that, for each DNN task, we can obtain $l_e+1$ training samples, where $l_e$ is $2$ based on the configuration of the shallow DNN in Fig. \ref{alexnet}. In contrast, without DT-assisted training data augmentation, very limited training samples can be obtained since only the workload evolution before the offloading of a DNN task can be used to construct training samples.

\begin{figure}
    \centering
    \includegraphics[width=5.5cm]{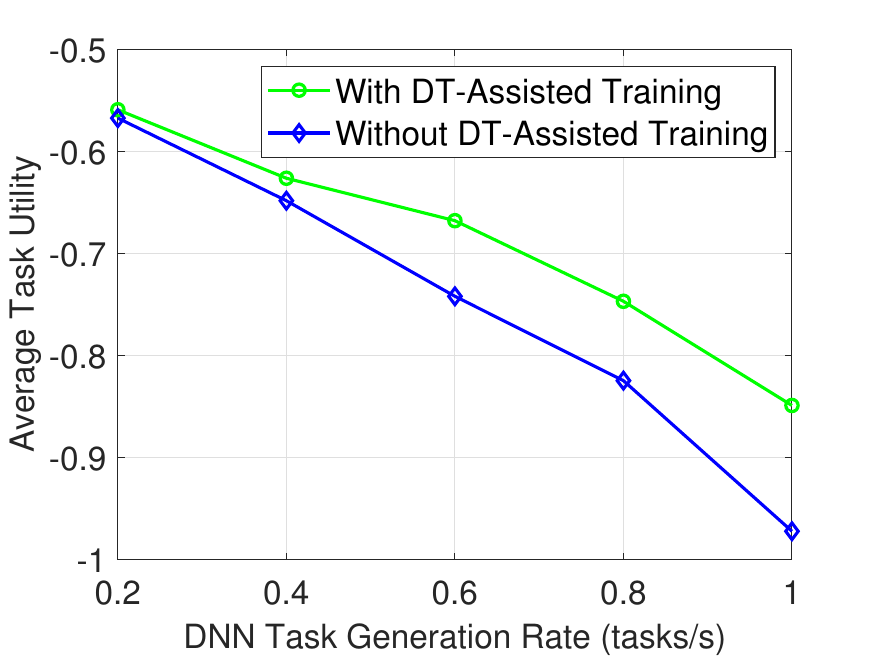}
    \caption{Average DNN task utility with and without DT-assisted training data augmentation.}
    \label{sim_fig3_ava_uti_versus_DNNrate_DT}
\end{figure}

\begin{figure}
    \centering
    \subfloat[]{\includegraphics[width=6.2cm]{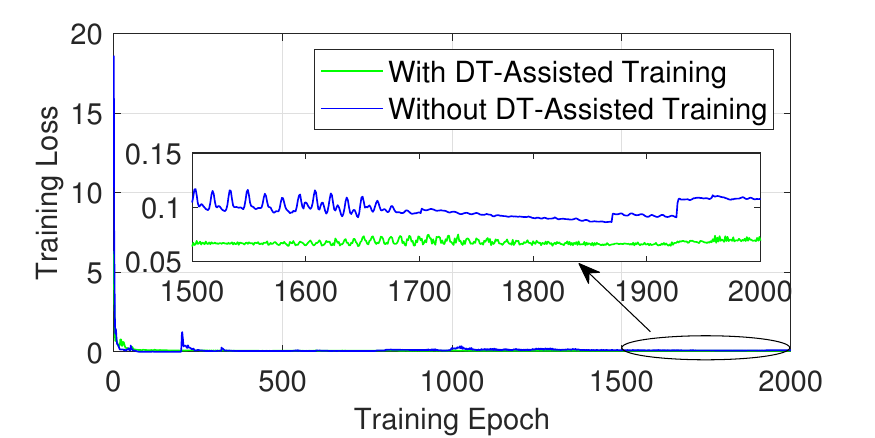}}\\
    \vspace*{-2pt}
    \subfloat[]{\includegraphics[width=6.2cm]{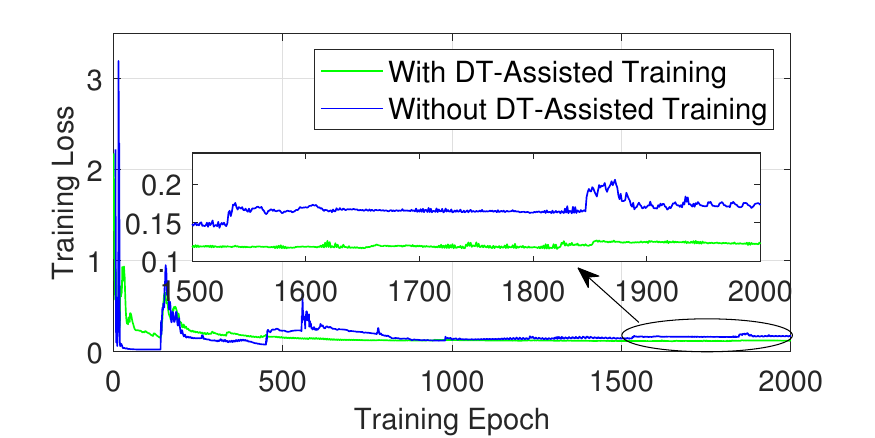}}
    \caption{Training loss with and without DT-assisted training data augmentation.}
    \label{sim_fig5_training_loss_versus_givenEdgeUtilization_epoch_DT}
\end{figure}

In Fig. \ref{sim_fig3_ava_uti_versus_DNNrate_DT}, the average DNN task utility versus the task generation rate, with and without DT-assisted training data augmentation, is shown, where the edge processing load is $0.9$. With DT-assisted training data augmentation, the average task utility increases. 
In addition, as the task generation rate increases, the performance gain increases. 
This is because the dynamics of the on-device workload increase with the task generation rate, which in turn requires a larger amount of and more diverse training data. 
The performance gain due to the DT-assisted data augmentation can also be explained by analyzing the training loss during the training of ContValueNet. In Fig. \ref{sim_fig5_training_loss_versus_givenEdgeUtilization_epoch_DT}, with the same simulation settings as those used to generate Fig. \ref{sim_fig6_training_samples_versus_epoch}, we can observe that without DT-assisted data augmentation, the training loss fluctuates more due to overfitting caused by insufficient training data. In contrast, with DT-assisted data augmentation, the training loss decreases more steadily to a lower level at the end of the training. 
\vspace*{3pt}

\subsection{Complexity Reduction by Decision Space Reduction}
\begin{figure}
    \centering
    \subfloat[]{\includegraphics[width=6.2cm]{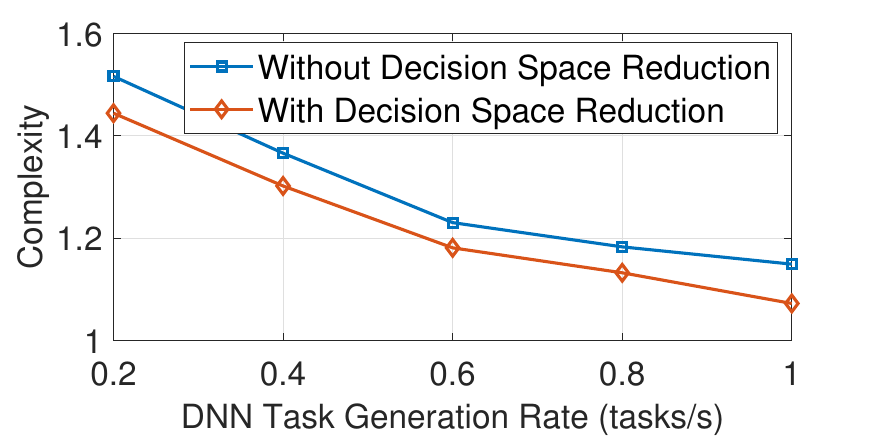}
    \label{sim_fig_reduction_mechanism_complexity}}\\
    \vspace*{-2pt}
    \subfloat[]{\includegraphics[width=6.2cm]{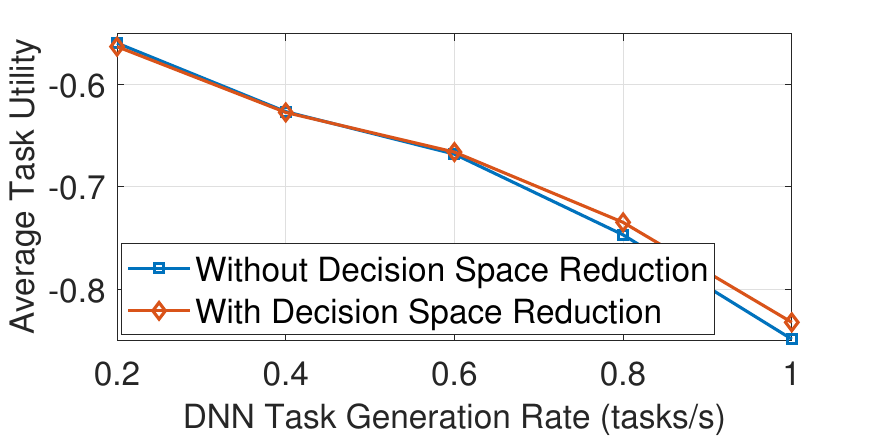}
    \label{sim_fig_reduction_mechanism_utility}}
    \vspace*{-6pt}
    \caption{(a) Complexity and (b) Average DNN task utility with and without decision space reduction.}
    \label{sim_fig_reduction_mechanism}
\end{figure}

In Fig. \ref{sim_fig_reduction_mechanism}, the performance of the proposed device-edge collaborative DNN inference with and without decision space reduction in Algorithm \ref{alg_1} is shown, where the edge processing load is $0.9$. As shown in Fig. \ref{sim_fig_reduction_mechanism_complexity}, with the decision space reduction, the proposed approach can be implemented with reduced complexity in terms of the average number of times when the controller determines whether to continue on-device inference for a DNN task based on (\ref{opt_stopping}). This is because with the decision space reduction, the on-device DNN inference continues if the corresponding offloading decision violates the necessary condition given in \textit{Lemma 1}. In addition, in Fig. \ref{sim_fig_reduction_mechanism_utility}, it can be observed that with the decision space reduction, the average task utility is nearly unaffected and even improved in the high regime of DNN task generation rate. This is because the necessary conditions sometimes prevent the controller from choosing non-optimal offloading decisions that could have been chosen due to the imperfect continuation value approximation by ContValueNet.

\section{Conclusion}\label{conclusion_chap}

In this paper, we have proposed a DT-assisted approach to device-edge collaboration on DNN inference, which can adapt to dynamic computing workloads at AIoT devices and edge servers with low signaling overhead. The proposed approach can be applied to support AIoT scenarios where densely deployed AIoT devices dynamically generate AI model inference tasks. For the future work, we will further explore the DTs to capture the time-varying data distribution for joint AI model training and AI model inference in AIoT.


\section*{Appendix A\\ Proof of Proposition 1}

\textit{Case 1:} If $T^{\rm lq}_{n-1}+\!T^{\rm lc}_{n-1}\le\Delta T_{n-1}$, the on-device queuing delay of the $n$-th DNN task $T^{\rm lq}_{n}$ equals to zero. Based on (\ref{queuing_delay_to}), $D_{n-1\to n}^{\rm lq} = \max\{\min{\{T^{\rm lq}_{n-1}\!-\!\begin{matrix}\Delta T_{n\!-\!1}\end{matrix}, 0\}} \!+ \!T_{n-1}^{\rm lc},0\}=0$. Similarly, we have $D_{m\to n}^{\rm lq}=0, \forall m< n$, we have $D_{m\to n}^{\rm lq}=0, \forall m\ge n$. Therefore, $T^{\rm lq}_{n}=\sum_{m=1}^{N} D_{m\to n}^{\rm lq}$.

\textit{Case 2:} If $T^{\rm lq}_{n-1}+\!T^{\rm lc}_{n-1}\!>\!\Delta T_{n-1}$, $T^{\rm lq}_{n} = T^{\rm lq}_{n-1}+\!T^{\rm lc}_{n-1}- \Delta T_{n-1}>0$, which indicates that one preceding DNN task (its index is denoted by $m_0$) is in the computing unit upon the generation of the $n$-th task. As a result, we have:
\begin{equation}
T_{m_0}^{\rm lq}\le\begin{matrix}\sum_{i=m_0}^{n-1}\Delta T_i\end{matrix}\le T_{m_0}^{\rm lq}+ T_{m_0}^{\rm lc},
\end{equation}
and for any $m$-th task, where $m_0+1\!\le\! m \!\le\! n-1$, $T_m^{\rm lq}\ge\sum_{i=m}^{n-1}\Delta T_i$.

Based on (\ref{queuing_delay_to}), $D_{m\to n}^{\rm lq}$ can be calculated by:
\begin{equation} \label{D_nm^lq}
D_{m\to n}^{\rm lq}= \left\{
\begin{aligned}
&T_{m}^{\rm lq}+ T_{m}^{\rm lc}- \begin{matrix}\sum_{i=m}^{n-1}\Delta T_i\end{matrix},~ m\!=\!m_0,
\\&T_{m}^{\rm lc},~m_0+1\!\le m \!\le\! n-1,
\\&0, \rm{otherwise}.
\end{aligned}
\right.
\end{equation}
Then, $\begin{matrix}\sum_{n=1}^{N} D_{m\to n}^{\rm lq}\end{matrix}$ can be calculated as:
\begin{equation}
\begin{aligned}
&\begin{matrix}\sum_{n=1}^{N} D_{m\to n}^{\rm lq}\end{matrix}
\\&= T_{m_0}^{\rm lq}+ T_{m_0}^{\rm lc} \!-\! \begin{matrix}\sum_{i=m_0}^{n-1}\Delta T_i\end{matrix}\! +\! \begin{matrix}\sum_{m=m_0+1}^{n-1}{T_{m}^{\rm lc}}\end{matrix}
\\&= (T_{m_0}^{\rm lq}+ T_{m_0}^{\rm lc} \!-\!\Delta T_{m_0})\!-\! \begin{matrix}\sum_{i=m_0+1}^{n-1}\Delta T_i\end{matrix}\! +\! \begin{matrix}\sum_{m=m_0+1}^{n-1}{T_{m}^{\rm lc}}\end{matrix}
\\&= T_{m_0+1}^{\rm lq}\!-\! \begin{matrix}\sum_{i=m_0+1}^{n-1}\Delta T_i\end{matrix}\! +\! \begin{matrix}\sum_{m=m_0+1}^{n-1}{T_{m}^{\rm lc}}\end{matrix}
\\&= T_{m_0+1}^{\rm lq}\!+\!T_{m_0+1}^{\rm lc}\!-\! \begin{matrix}\sum_{i=m_0+1}^{n-1}\Delta T_i\end{matrix}\! +\! \begin{matrix}\sum_{m=m_0+2}^{n-1}{T_{m}^{\rm lc}}\end{matrix}
\\&\cdots
\\&= T_{n-1}^{\rm lq}\!+\!T_{n-1}^{\rm lc} \!-\! \Delta T_{n-1}
\\&= T_{n}^{\rm lq}.
\end{aligned}
\end{equation}
\begin{figure*}[!t]
\normalsize
\begin{equation}\label{sum_D_nmlq}
\begin{split}
D_n^{\rm lq}&=\begin{matrix}\sum_{m=1}^{N}\end{matrix} D_{n\to m}^{\rm lq}
\\&= q_0 T_n^{\rm lc}+\begin{matrix}\sum_{m=n+q_0+1}^{n+q_1}\end{matrix}(T_n^{\rm lq}+T_n^{\rm lc}-\begin{matrix}\sum_{i=n}^{m-1}\end{matrix}\Delta T_i)
\\ &=(q_1-q_0)T_n^{\rm lq}+q_1T^{\rm lc}_n-(q_1-q_0)\begin{matrix}\sum_{i=n}^{n+q_0}\end{matrix}\Delta T_i -\begin{matrix}\sum_{i=q_0+n+1}^{q_1+n-1}\end{matrix}(q_1-i+n)\Delta T_i
\end{split}
\end{equation}
\begin{equation}\label{sum_Q}
\begin{split}
\begin{matrix}\sum_{t=t_{n,0}}^{t_{n,0}+(T_n^{\rm lc}/\Delta T)-1}  \end{matrix}Q^{D}(t)\Delta T&= \begin{matrix}q_0(-t_{n,0}\Delta T\!+\!\sum_{i=1}^{n+q_0} \Delta T_i)\!+\!\sum_{m=q_0+n+1}^{q_1+n-1}\end{matrix}(m\!-\!n)\Delta T_m \!+\!q_1(t_{n,0}\Delta T\!+\!T_{n}^{\rm lc}-\begin{matrix}\sum_{i=1}^{n+q_1-1}\Delta T_i)\end{matrix}
\\&=q_0(-T_n^{\rm lq}\!+\!\begin{matrix}\sum_{i=n}^{n+q_0}\end{matrix}\Delta T_i)+\begin{matrix}\sum_{m=q_0+n+1}^{q_1+n-1}\end{matrix}(m\!-\!n)\Delta T_m\! + \!q_1(T_n^{\rm lq}\!+\!T_n^{\rm lc}\!-\!\begin{matrix}\sum_{i=n}^{n+q_1-1}\end{matrix}\Delta T_i)
\\&=(q_1\!-\!q_0)T_n^{\rm lq}+q_1T^{\rm lc}_n\!-\!(q_1\!-\!q_0)\begin{matrix}\sum_{i=n}^{n+q_0}\end{matrix}\Delta T_i\!+\! (\begin{matrix}\sum_{m=q_0+n+1}^{q_1+n-1}\end{matrix}(m\!-\!n)\Delta T_m\!-\!q_1\begin{matrix}\sum_{i=q_0+n+1}^{q_1+n-1}\end{matrix}\Delta T_i)
\\&=(q_1\!-\!q_0)T_n^{\rm lq}\!+\!q_1T^{\rm lc}_n\!-\!(q_1\!-\!q_0)\begin{matrix}\sum_{i=n}^{n+q_0}\end{matrix}\Delta T_i
-\begin{matrix}\sum_{i=q_0+n+1}^{q_1+n-1}\end{matrix}(q_1-i+n)\Delta T_i
\end{split}
\end{equation}
\hrulefill
\vspace*{4pt}
\end{figure*}

\section*{Appendix B\\ Proof of Proposition 2}
When $T_n^{\rm lc}=0$, $D_{n\to m}^{\rm lq}=0, \forall 1 \le m\le N$. As a result, $D_n^{\rm lq}=\begin{matrix}\sum_{m=1}^{N}\end{matrix} D_{n\to m}^{\rm lq}=0$. When $T_n^{\rm lc}\ge\Delta T$, for brevity, we denote $Q^{D}(t_{n,0})$ and $Q^{D}(t_{n,0}+(T_n^{\rm lc}/\Delta T)-1)$ by $q_0$ and $q_1$, which represent the number of tasks in the on-device task queue in the time slots when the on-device inference for the $n$-th DNN task starts and completes, respectively. During the on-device inference for the $n$-th task, the $m$-th task is in the on-device task queue, where $n\!< m\! \le n+q_1$. We can derive the queuing delay of the $m$-th task that results from the on-device inference of the $n$-th task as:
\begin{equation} \label{D_nm^lq}
D_{n\to m}^{\rm lq}=\left\{
\begin{aligned}
&T_n^{\rm lc},~ n\!<\! m\!\le\! n\!+\!q_0,
\\&T_n^{\rm lq}\! +\! T_n^{\rm lc}-\begin{matrix}\sum_{i=n}^{m-1}\end{matrix}\Delta T_i, ~n\!+\!q_0\!+\!1\!\le\! m\! \le \!n\!+\!q_1,
\end{aligned}
\right.
\end{equation}
and calculate on-device workload during the on-device inference of the $n$-th task by:
\begin{equation} \label{Qlocal(t)}
Q^{D}(t)=\left\{
\begin{aligned}
& q_0, ~~~~~~~~~~~~~~~~~~~t_{n,0}\le t < \frac{1}{\Delta T}\begin{matrix}\sum_{i=1}^{n+q_0}\end{matrix}\Delta T_i,
\\&\!m\!-\!n,~ \frac{1}{\Delta T}\begin{matrix}\sum_{i=1}^{n+q_0}\end{matrix}\Delta T_i \le t< \frac{1}{\Delta T}\begin{matrix}\sum_{i=1}^{m-1}\end{matrix}\Delta T_i,
\\&q_1, ~~\frac{1}{\Delta T}\begin{matrix}\sum_{i=1}^{n+q_1-1}\end{matrix}\Delta T_i \le t< t_{n,0}+\frac{1}{\Delta T}T_n^{\rm lc},
\end{aligned}
\right.
\end{equation}
where $q_0+n\!\le m\! \le q_1+n$. Based on (\ref{D_nm^lq}) and (\ref{Qlocal(t)}), we can calculate $\sum_{m=1}^{N} D_{n\to m}^{\rm lq}$ and $\begin{matrix}\sum_{t=t_{n,0}}^{t_{n,0}+(T_n^{\rm lc}/\Delta T)-1} Q^{D}(t)\Delta T\end{matrix}$ using (\ref{sum_D_nmlq}) and (\ref{sum_Q}), respectively, and it can be observed from (\ref{sum_D_nmlq}) and (\ref{sum_Q}) that they are equal.

\bibliographystyle{IEEEtran}
\bibliography{Hss_Ref}

\end{document}